\documentclass[]{aa}
\usepackage{txfonts}
\usepackage{natbib}
\usepackage{graphicx}
\usepackage{longtable}
\bibpunct{(}{)}{;}{a}{}{,}
\begin{document}

\title{Spectroscopy of horizontal branch stars in $\omega$\,Centauri
\thanks{Based on observations with the ESO Very Large Telescope at Paranal Observatory, Chile (proposal ID 076.D-0810)}
\thanks{Table~2 is only available in electronic form at the CDS via anonymous ftp to cdsarc.u-strasbg.fr (130.79.128.5)
or via http://cdsweb.u-strasbg.fr/cgi-bin/qcat?J/A+A/}
}

\author{
C. Moni Bidin \inst{1}
\and
S. Villanova \inst{1}
\and
G. Piotto \inst{2,3}
\and
S. Moehler \inst{4}
\and
S. Cassisi \inst{5}
\and
Y. Momany \inst{3,6}
}

\institute{
Departamento de Astronom\'ia, Universidad de Concepci\'on, Casilla 160-C, Concepci\'on, Chile
\and
Dipartimento di Fisica e Astronomia ``Galileo Galilei", Universit\`a di Padova, Vicolo dell'Osservatorio 3, I-35122 Padova, Italy
\and
INAF, Osservatorio Astronomico di Padova, Vicolo  dell'Osservatorio 5, 35122, Padova, Italy
\and
European Southern Observatory, Karl-Schwarzschild-Str. 2. 85784 Garching, Germany
\and
INAF Osservatorio Astronomico di Collurania, via Mentore Maggini, 64100 Teramo, Italy
\and
European Southern Observatory, Alonso de Cordova 3107, Vitacura, Santiago, Chile
}
\date{Received / Accepted }


\abstract
{}
{We analyze the reddening, surface helium abundance and spectroscopic mass of 115 blue horizontal branch (HB) and
blue hook (BH) stars in $\omega$\,Centauri, spanning the cluster HB from the blue edge of the instability strip
($T_\mathrm{eff}$=8\,000~K) to BH objects with $T_\mathrm{eff}\approx$50\,000~K.}
{The temperatures, gravities, and surface helium abundances were measured on low-resolution spectra fitting the Balmer
and helium lines with a grid of synthetic spectra. From these parameters, the mass and reddening were estimated.}
{The mean cluster reddening is $E(B-V$)=0.115$\pm$0.004, in good agreement with previous estimates, but we evidence
a pattern of differential reddening in the cluster area. The stars in the western half are more reddened than in the
southwest quadrant by 0.03--0.04~magnitudes. We find that the helium abundances measured on low-resolution spectra are
systematically lower by 0.20-0.25~dex than the measurements based on higher resolution. No difference in surface
helium abundance is detected between HB stars in $\omega$\,Centauri and in three comparison clusters, and the stars
in the range 11\,500--20\,000~K follow a trend with temperature, which probably reflects a variable efficiency of the
diffusion processes. There is mild evidence that two families of extreme HB (EHB) cluster stars
($T_\mathrm{eff}\geq$20\,000~K) could exist, as observed in the field, with $\sim$15\% of the objects being
helium depleted by a factor of ten with respect to the main population. The distribution of helium abundance above
30\,000~K is bimodal, but we detect a fraction of He-poor objects lower than previous investigations. The observations
are consistent with these being stars evolving off the HB. Their spatial distribution is not uniform across the cluster,
but this asymmetric distribution is only marginally significative. We also find that EHB stars with anomalously high
spectroscopic mass could be present in $\omega$\,Centauri, as previously found in other clusters. The derived
temperature-color relation reveals that the HB stars hotter than $\sim$11\,000~K are fainter than the expectations of
the canonical models in the $U$ band, while no anomaly is detected in $B$ and $V$. This behavior, not observed in
NGC\,6752, is a new peculiarity of $\omega$\,Centauri HB stars. More investigation is needed to reach a full
comprehension of this complex observational picture.}
{}


\authorrunning{Moni Bidin et al.}
\mail{cmbidin@astro-udec.cl}
\titlerunning{}
\maketitle


\section{Introduction}
\label{s_intro}

Horizontal branch (HB) stars in Galactic globular clusters (GCs) are old stars of low initial mass (0.7--0.9 $M_{\sun}$)
that, after the exhaustion of hydrogen in the stellar core and the ascension along the red giant branch, finally
ignite He burning in the core \citep{Hoyle55,Faulkner66}. Despite this general comprehension, our knowledge of cluster
HB stars still presents many grey areas \citep[see][for a recent review]{Catelan09}. In particular, recent
observations of HB stars hotter than 20\,000~K (extreme HB stars, EHB) in GCs have left many questions waiting for a
proper answer \citep{Moni09b}.

The deep morphological differences observed among the HBs of the Galactic GCs are not fully explained because the
cluster metallicity alone cannot account for them \citep{Sandage67,VanDenBergh67}. The HB morphology has been linked,
among others, to cluster age \citep{Dotter10}, cluster concentration \citep{FusiPecci93}, stellar rotation
\citep{Peterson83}, cluster mass \citep{RecioBlanco06}, and the environment of formation \citep{FraixBurnet09}. However,
none of the proposed second parameters could satisfactorily reproduce the complex observed behavior.
\citet{Gratton10} showed that a set of at least three parameters is required to describe the observations. The
recent discovery that many clusters host stellar sub-populations with different helium content \citep{Piotto05,Piotto07}
has given new strength to the proposition that helium abundance could be a key parameter governing the cluster HB
morphology \citep{DAntona02,DAntona05,Lee05}. In this scenario, the blue HB stars observed in many GCs would be the
progeny of the He-enriched second stellar generation. Unfortunately, diffusion processes are active in the
atmosphere of HB stars hotter than $\sim$11\,500~K \citep{Michaud83,Michaud08,Quievy09}, causing photometric anomalies
\citep{Grundahl99} and deep alteration of the surface chemical composition \citep{Behr03}. As a consequence, the
direct measurement of the primordial helium abundance is possible only for blue HB stars cooler than about 11\,500~K
\citep{Villanova09,Villanova12}. \citet[][hereafter Paper~I]{Moni11a} recently searched for indirect evidence of
helium enrichment among blue HB stars in $\omega$\,Centauri, a cluster known to host a very complex mix of at least
six sub-populations \citep{Bellini10}. \defcitealias{Moni11a}{Paper~I} Their results are surprisingly puzzling. In
brief, the measured gravities are systematically lower than the predictions of canonical evolutionary models with
solar helium abundance, in agreement with the expectations for He-enriched models. This behavior was not observed in
three other clusters previously analyzed, and it is unique of $\omega$\,Cen stars. However, the calculated masses are
unrealistically low, and the low gravities can thus not be straightforwardly interpreted as a direct evidence of
helium enrichment.

The UV color-magnitude diagram (CMD) of the most massive GCs has revealed an additional puzzling feature, the so-called
blue hook (BH), a population of stars bluer than the canonical end of the HB
\citep{Whitney98,Piotto99,DCruz00,Brown01,Rosenberg04,Momany04,Busso07,Ripepi07}. The formation of these extremely hot
objects ($T_\mathrm{eff}\geq$32\,000~K) cannot be explained by the canonical stellar evolution theories. They were
proposed to be the progeny of stars that, due to an unusually large mass loss, left the red giant branch before the
helium flash and ignited helium on the white dwarf cooling sequence \citep{Castellani93,DCruz96,Brown01}. Due to
the very low efficiency of the H-burning in the shell, and hence the very low entropy barrier present at the shell
location, these stars can experience a He flash-induced mixing inside the He core able to reach the H-rich envelope
\citep{Sweigart97,Cassisi03}. As a consequence of this process, some amount of H can be dredged down the stellar interior,
and He-burning products can be dredged up the stellar surface. Depending on the efficiency of this He flash-induced mixing,
some hydrogen can remain in the envelope \citep{Lanz04}, while the surface carbon abundance is increased to 1-5\% by mass.
On the other hand, \citet{Lee05} suggested that BH stars in $\omega$\,Cen are the progeny of the helium-enriched main
sequence population. \citet{DAntona10} proposed that extra-mixing processes during the red giant phase can increase their
surface helium abundance up to $Y\approx$0.8, required to explain their photometric behavior. Their carbon abundance should,
however, not be enhanced. The high carbon abundances found by \citet{Moehler11} pose a problem to these scenarios, as they
indicate that some extra process (beyond helium enrichment) is required to explain the BH stars. Moreover, while the
majority of their targets above 30\,000~K showed a solar or super-solar surface helium abundance, they detected a
sub-population of very hot helium-poor EHB stars, which they propose are post-HB stars evolving toward the white dwarf
cooling sequence.

We measured the surface parameters of a large sample of HB and BH stars in $\omega$\,Cen to gather new information
about their properties. The results for the temperature, gravities, and masses of the stars with
$T_\mathrm{eff}\leq$32\,000~K were presented in \citetalias{Moni11a}. In this paper, we will present the results on
the surface helium abundance for all the stars, focusing on the BH candidates hotter than 32\,000~K, which were
not analyzed in \citetalias{Moni11a}. Whenever applicable, we will compare our results to those of \citet{Moni07,Moni09},
who measured the parameters of HB stars in NGC\,6752, M\,80, and NGC\,5986. At variance with $\omega$\,Cen, these three
clusters do not show main sequence (MS) splitting, although the MS of NGC\,6752 is broadened \citep{Milone10}.
\citet{Moni07,Moni09} used the same instrument, software, and models as we did, and the comparison can thus easily reveal
intrinsic differences between the clusters.


\section{Observations and data reduction}
\label{s_obs}

\begin{figure}
\begin{center}
\includegraphics[width=9cm]{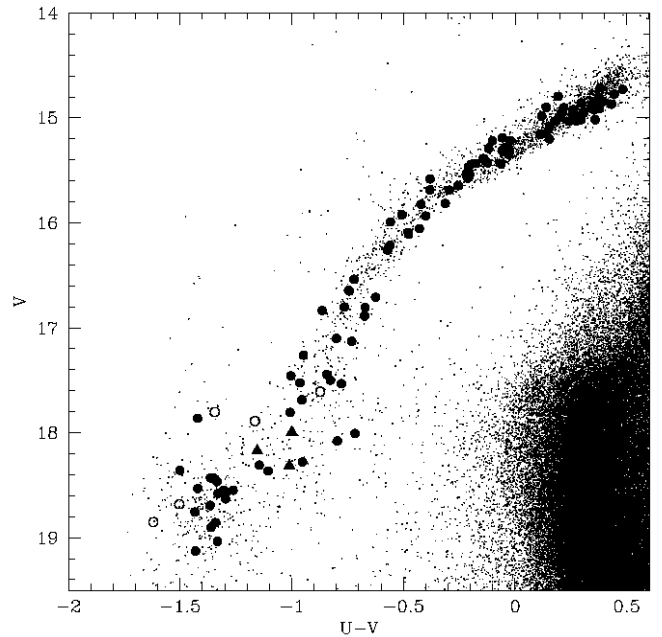}
\caption{Position of the target stars in the cluster CMD. The empty circles indicate hot 
($T_\mathrm{eff}\geq$32\,000~K) helium-poor stars. Full triangles show the objects with anomalously high
spectroscopic mass.}
\label{f_cmd}
\end{center}
\end{figure}

Our observations targeted 115 HB stars in $\omega$\,Cen, selected from the optical photometry of \citet{Bellini09}. They
span a wide range of the cluster HB, from the blue edge of the RR Lyrae gap ($T_\mathrm{eff}\sim$7\,800~K) to BH
candidates at $T_\mathrm{eff}\geq$32\,000~K. The distribution of the targets in the cluster CMD is shown in Fig.~\ref{f_cmd}.

The data were collected at the Paranal Observatory in service mode, with the FORS2 spectrograph mounted on the UT1
telescope. The observations were performed under a variety of sky conditions, as shown in the log of the observations given
in Table~\ref{t_log}. The instrument was used in multi-slit (MXU) mode, collecting between 12 and 24 spectra
in each of the seven masks that were employed. The slit width of $0\farcs 5$ and the 600B grism returned a spectral
resolution R$\approx$1600 in the range 3\,450--5\,900~\AA. Two 45-minutes exposures were acquired for the masks comprising
only bright stars, while three similar exposures were collected when faint stars were involved. The final spectra thus had
a signal-to-noise ratio between 40 and 180, although the targets span a range of about four magnitudes.

\begin{figure*}
\begin{center}
\includegraphics[width=16cm]{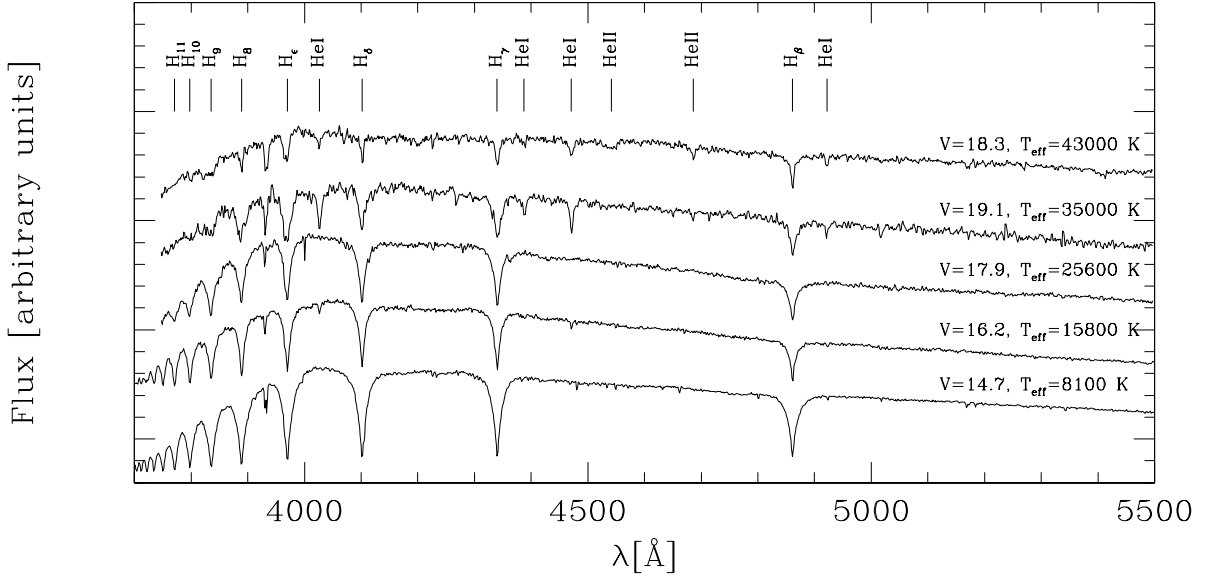}
\caption{Example of the flux-calibrated extracted spectra of stars at different temperatures. The hydrogen and helium lines
used in the fitting routine are also indicated.}
\label{f_spectra}
\end{center}
\end{figure*}

The spectra were de-biased, flat-fielded, and wavelength-calibrated with the FORS
pipeline\footnote{\small{www.eso.org/sci/data-processing/software/pipelines/index.html}}. The accuracy of the wavelength
calibration was $\sim$5~km~s$^{-1}$. The spectra were then extracted with standard IRAF\footnote{\small{IRAF is distributed
by the National Optical Astronomy Observatories, which are operated by the Association of Universities for Research in
Astronomy, Inc., in cooperative agreement with the National Science Foundation.}} routines. They were corrected
subtracting the sky background within the same slit, whose minimum length was $6\arcsec$, and then flux-calibrated. The
response curve was obtained from the spectrum of the standard star LTT4816 \citep{Hamuy92}, collected during the
observations. Some examples of the reduced spectra are shown in Fig.~\ref{f_spectra}.

\begin{table}
\caption{Log of the observations.}
\label{t_log}
\begin{tabular}{lllll}
\hline
\hline
Start of & Airmass & seeing  & \multicolumn{2}{c}{Moon} \\
observation & & & distance & illumi-\\
(UT) & & ($\arcsec$) & ($\deg$) & nation\\
\hline
2006-01-19T07:34:14 & 1.25 & 1.0 &  58 & 0.80\\
2006-01-19T08:02:00 & 1.19 & 0.9 &  57 & 0.80\\
2006-01-19T08:29:23 & 1.15 & 1.1 &  57 & 0.80\\
2006-02-09T07:42:57 & 1.11 & 0.7 & 120 & 0.86\\
2006-02-09T08:15:37 & 1.09 & 0.8 & 120 & 0.86\\
2006-02-09T08:49:06 & 1.09 & 0.7 & 119 & 0.86\\
2006-02-14T05:23:28 & 1.35 & 2.2 &  71 & 0.99\\
2006-03-23T08:47:02 & 1.30 & 0.6 &  63 & 0.43\\
2006-03-23T09:00:09 & 1.34 & 0.5 &  63 & 0.43\\
2006-03-31T07:54:08 & 1.27 & 1.5 & 148 & 0.05\\
2006-04-01T06:24:15 & 1.12 & 0.8 & 147 & 0.11\\
2006-04-01T06:55:04 & 1.15 & 0.7 & 147 & 0.11\\
\hline
\end{tabular}
\end{table}

\begin{figure}
\begin{center}
\includegraphics[width=9cm]{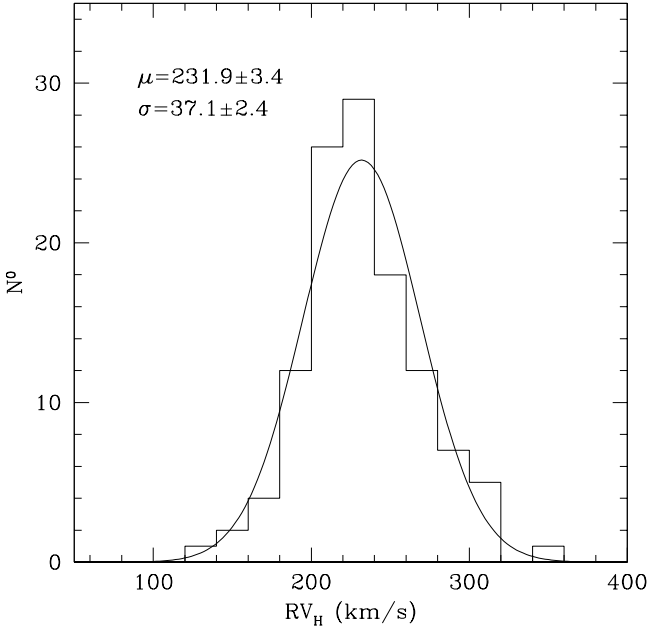}
\caption{Histogram of the distribution of the measured heliocentric RVs. A Gaussian fit with the given
mean and dispersion is also indicated.}
\label{f_rvs}
\end{center}
\end{figure}

The heliocentric radial velocity (RV) of the target stars was measured with the IRAF task {\it fxcor}. Each spectrum
was cross-correlated \citep{Tonry79} with a synthetic template with temperature and gravity similar to those of the target,
as deduced from its position on the cluster HB. Previous investigations have shown that the RV measurements are negligibly
affected by the exact choice of the template \citep{Morse91,Moni11b}. The results are given in the tenth column of
Table~\ref{t_res}, and the distribution of the RVs is shown in Fig.~\ref{f_rvs}. The error deriving from the
cross-correlation procedure is $\sim$30~km~s$^{-1}$. The mean RV of the sample is
$\overline{\mathrm{RV}}=231.9\pm3.4$~km~s$^{-1}$, in excellent agreement with the cluster RV of 232.1~km~s$^{-1}$ quoted
by \citet[][2010 December Web version\footnote{http://physwww.physics.mcmaster.ca/\%7Eharris/mwgc.dat}]{Harris96}. The
velocities follow a Gaussian distribution with no evident outliers, and they are therefore compatible with the assumption
that all the targets are cluster members. However, the observed dispersion is
37.1$\pm$2.4~km~s$^{-1}$, larger than that expected from the observational errors and the internal cluster dispersion
\citep[$\sim$13~km~s$^{-1}$,][]{Sollima05}. In fact, quadratically subtracting the estimated wavelength calibration and
measurement errors, we obtain an intrinsic dispersion of $\sim$21~km~s$^{-1}$, incompatible with the cluster internal
dispersion at any distance from the center \citep{Scarpa10}. This most probably indicates that the imperfect centering of
the targets inside the slits introduced an additional error of the order of $\sim$20~km~s$^{-1}$, i.e., about one
tenth of resolution element, or 0.4 pixels. This effect is frequent in multi-object slit spectroscopy
\citep[see the analysis of][]{Moni06}.


\section{Measurements}
\label{s_meas}

The temperature, gravity, and atmospheric helium abundance of the target stars were measured by fitting the observed
hydrogen and helium lines with synthetic spectra. The stars at the coolest end of our sample
($T_\mathrm{eff}\leq$12\,000--13\,000~K) were fitted with a grid of model spectra computed with Lemke's
version\footnote{\small{http://a400.sternwarte.uni-erlangen.de/\~{}ai26/linfit/linfor.html}} of
the LINFOR program (developed originally by Holweger, Steffen, and Steenbock at Kiel University), fed with local
thermodynamic equilibrium (LTE) model atmospheres of cluster metallicity ([M/H]=$-$1.5) computed with ATLAS9
\citep{Kurucz93}. The helium abundance was kept
fixed to solar value, as expected for stars not affected by diffusion processes, because the He lines of cool stars
are weak and not observed at our resolution. Stars hotter than 13\,000~K, as deduced from their position in the CMD, or
showing evidence of active atmospheric diffusion, i.e., strong iron lines between 4450 and 4600~\AA\ \citep{Moehler99},
were fitted with models of super-solar metallicity ([M/H]=+0.5) and variable surface helium abundance to account for
the effects of radiative levitation of heavy elements \citep{Moehler00}. This was done even for five warm stars
not fully satisfying these criteria (stars \#82876, \#133061, \#75469, \#100288, and \#100817, with
$T_\mathrm{eff}$=11\,500--13\,000~K), because the helium lines of the model with solar helium abundance were too strong
compared to the observed ones. The helium abundance was thus determined during the fitting routine. Stars bluer than the
canonical end of the EHB in the CMD ($T_\mathrm{eff}\geq$32\,000~K) were fitted with the grid of metal-free non-LTE
models described in \citet{Moehler04}, calculated as in \citet{Napiwotzki97}. We found that the use of these models did
not improve the quality of the fit (expressed by the $\chi^2$ statistics) for stars between 30\,000 and 32\,000~K. When
the routine indicated a helium abundance next to solar ($\log{(N_\mathrm{He}/N_\mathrm{tot})}=-1$) or higher, the fit was
repeated with the helium-rich non-LTE models, calculated with a modified version of the code of \citet{Werner99}, with
the model atoms of \citet{Werner96}. The star \#178139 was fitted with helium-poor models, despite its solar helium
abundance, because the use of the He-rich models degraded the quality of the fit noticeably.

\begin{figure}
\begin{center}
\includegraphics[width=9cm]{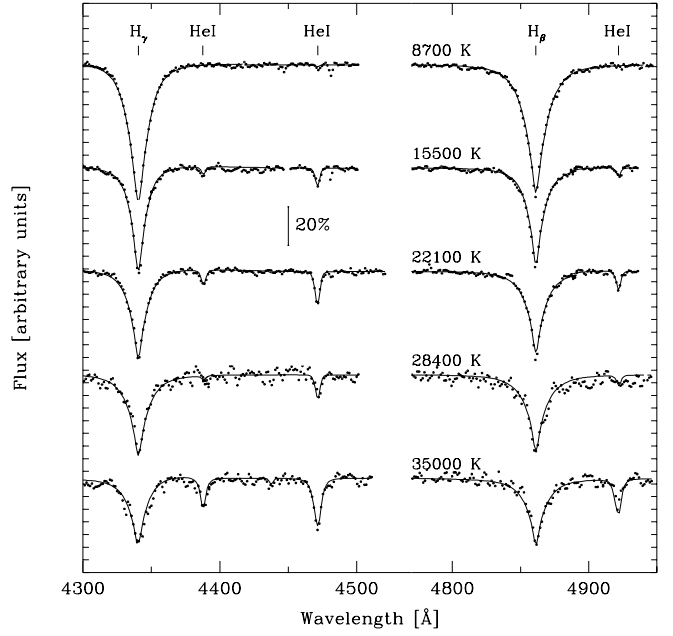}
\caption{Example of fit of the most prominent spectral features. The observed spectra are shown with circles, the
best-fit synthetic spectrum is indicated by the full curve.}
\label{f_fitsyn}
\end{center}
\end{figure}

The routines developed by \citet{Bergeron92} and \citet{Saffer94}, as modified by \citet{Napiwotzki99}, were used to
derive the stellar parameters. They normalize both the model and observed spectra using the same points for the
continuum definition and employ a $\chi^2$ test to establish the best fit. The noise in the continuum spectral
regions is used to estimate the $\sigma$ for the calculation of the $\chi^2$ statistics, which the routines use to
estimate the errors on the parameters \citep[see][]{Moehler99}. However, they thus neglect other sources of errors,
such as those introduced by the normalization procedure, the sky subtraction, and the flat-fielding. Therefore, the
resulting uncertainties were multiplied by three to obtain a more realistic estimate of the true errors (R. Napiwotzki
2005, private communication). The lines used in the fitting procedure included the Balmer series from H$_\beta$ to
H$_{12}$, except the H$_\epsilon$ to avoid the blended $\ion{Ca}{II}$~H line, and four $\ion{He}{I}$ lines (4026~\AA,
4388~\AA, 4471~\AA, 4922~\AA) for the stars whose helium abundance was a free fit parameter. Two $\ion{He}{II}$ lines
(4542~\AA, 4686~\AA) were also used, when visible, in the spectra of the hottest stars. Some fits of the observed
spectral features are shown in Fig.~\ref{f_fitsyn} as an example for two Balmer and three $\ion{He}{I}$ lines of
five stars at different temperatures.

Stellar masses were estimated from the derived stellar parameters through the relation
\begin{equation}
\log{\frac{M}{M_{\sun}}}=\log{\frac{g}{g_{\sun}}}-4\cdot \log{\frac{T_{eff}}{T_{eff,\sun}}}+\log{\frac{L}{L_{\sun}}},
\label{eq_mass1}
\end{equation}
where
\begin{equation}
\log{\frac{L}{L_{\sun}}}=-0.4\cdot(V - (m-M)_0 - 3.1\cdot E(B-V) + BC_V - M_\mathrm{bol,\sun}).
\label{eq_mass2}
\end{equation}
We assumed $T_{\sun}$=5777 K, $\log{g_{\sun}}$=4.44, $(m-M)_0=13.75\pm 0.13$ \citep{Vandeven06}, and
$E(B-V)=0.12\pm$0.01 \citep[][December 2010 Web version]{Harris96}. We adopted this reddening for all the
stars instead of the value spectroscopically derived for each target (Sect.~\ref{ss_reddening}), because our
results are scattered around this mean reddening, with negligible trend with temperature. Hence, the use of the
individual reddening only adds noise without altering the general trend. The bolometric correction (BC$_V$) was
derived from the effective temperature through the empirical calibration of \citet{Flower96}. We therefore
fixed $M_{bol,\sun}$=4.75 because $M_{V,\sun}$=4.83 \citep{Binney98}, and the \citet{Flower96}
BC$_V$--$T_\mathrm{eff}$ relation returns BC$_{\sun}=-$0.08. Errors on masses were derived from propagation of
errors. The resulting temperature, gravity, helium abundance, and mass of each target are given in
Table~\ref{t_res}.


\section{Results}
\label{s_res}

\subsection{Comparison with Moehler et al. (2011)}
\label{ss_Moehler}

\begin{figure}
\begin{center}
\includegraphics[width=9cm]{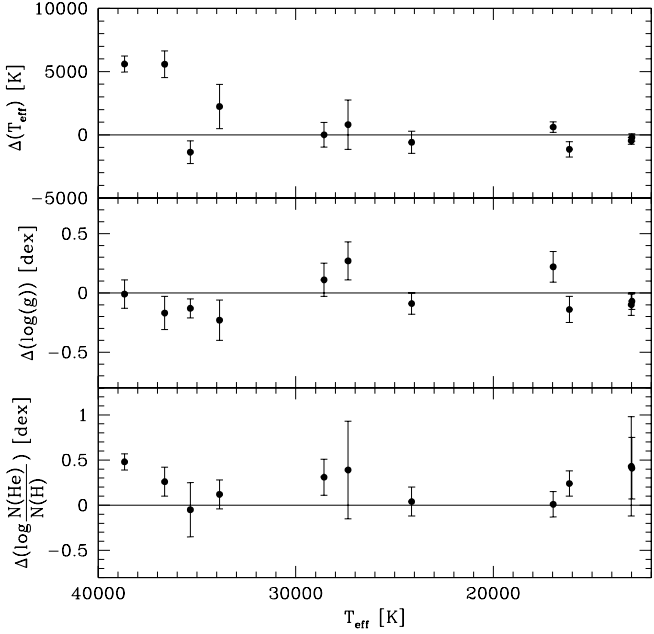}
\caption{Comparison between our results and those of \citet{Moehler11} for the 11 stars in common. The differences
(in the sense ours$-$Moehler et al.'s) in temperature (upper panel), gravity (central panel), and helium abundance
(lower panel) as a function of temperature are shown.}
\label{f_Moehler}
\end{center}
\end{figure}

Our sample comprises 11 stars studied by \citet{Moehler11}. They are indicated with an ``M'' in the last column
of Table~\ref{t_res}. Moehler et al. measured the stellar parameters with the same procedure and software used by us,
but with high-resolution spectra of shorter wavelength coverage. They also employed the same models for stars with
$T_\mathrm{eff}\leq$20\,000~K, while for hotter stars they used metal-poor ([M/H]=$-$1.5) non-LTE model atmospheres.
The comparison for the stars in common is shown in Fig.~\ref{f_Moehler}, where the errors bars indicate the quadratic
sum of the uncertainties in the two works. As commented in \citetalias{Moni11a}, the temperature and gravity measured
in the two investigations for the seven stars in common cooler than 32\,000~K agree well: the mean differences
(ours$-$Moehler et al.'s) are only $\Delta(T_\mathrm{eff})=-$143~K and $\Delta(\log{g})$=0.03~dex. Our mass estimates
are on average higher by 0.08~$M_{\sun}$, as a consequence of the fainter magnitudes of the \citet{Castellani07}
catalog, adopted by \citet{Moehler11}. On the contrary, the four hottest stars suggest that some systematic could be
present between the BH stars. In fact, our temperatures are on average higher by $\sim$3\,000~K, and our gravities (and
masses) lower by $\sim$0.14~dex. The use of different models is expected to cause such effect (S. Dreizler 2012, priv.
comm.), but the stars in common are too few and the detected offsets could not be significative. For example, the
difference in temperature is entirely due to the two hottest stars only.

\begin{figure}
\begin{center}
\includegraphics[width=9cm]{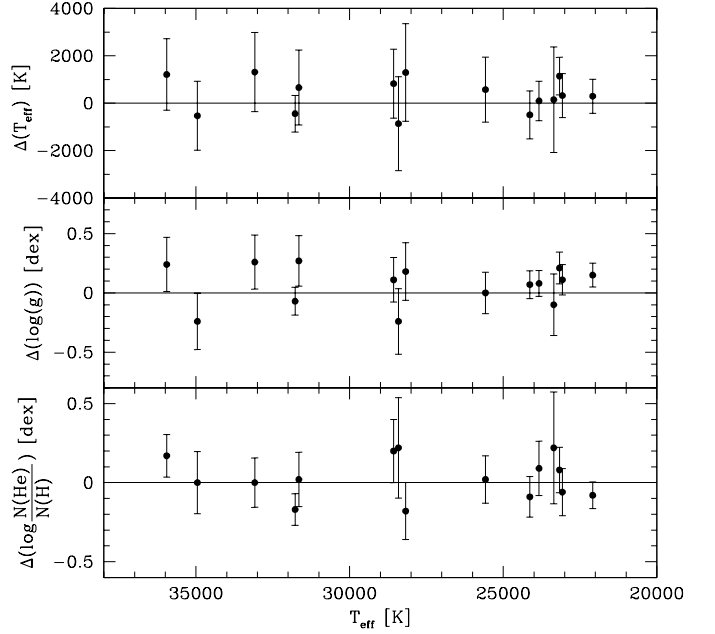}
\caption{Same as Fig.~\ref{f_Moehler}, but comparing our results with those derived by analysing our data using the
model spectra of \citet{Moehler11}.}
\label{f_newMoehler}
\end{center}
\end{figure}

To investigate further the systematics introduced by the use of different models, we re-fitted a subset of 15
stars with $T_\mathrm{eff}\geq$20\,000~K (indicated with ``T'' in the last column of Table~\ref{t_res}) with the
same models used by \citet{Moehler11}, calculated with TLUSTY\footnote{http://nova.astro.umd.edu} \citep{Hubeny95}.
More details about the models atoms and the atomic data can be found in \citet{Lanz03,Lanz07} and \citet{Moehler11}.
The comparison between the parameters obtained with our
models and these non-LTE ones (in the sense our models$-$TLUSTY) is shown in Fig.~\ref{f_newMoehler}, where the
error bars indicate the quadratic sum of the errors in the two measurements. The comparison only partially explains
the differences with \citet{Moehler11} results. The temperatures measured with our models are higher, but by only
$\sim$370~K on average, much less than the differences seen at the hotter end of Fig.~\ref{f_Moehler}. Our gravities
are higher, and not lower, by 0.07~dex, which is not a significative offset compared to the errors. In both cases, no
trend with temperature is observed. We also note that the differences in temperature and gravity are strongly
correlated. As an additional test, we repeated the measurements on the \citet{Moehler11} FLAMES spectra, after
degrading their resolution to match that of FORS data. Very similar results were found, with no evidence of a
significant offset in temperature or gravity. In conclusion, the use of different models accounts only for a small
offset in temperature. The differences with Moehler et al.'s results could be an effect of the small quantity of hot
stars in common. However, other evidences later in our analysis (Sections~\ref{ss_hehot} and~\ref{ss_mass}) suggest
that an offset could indeed be present.

Figure~\ref{f_Moehler} shows an offset in the surface helium abundances, our measurements being systematically
higher than those of \citet{Moehler11} by 0.24~dex on average. No trend with temperature is visible. This offset is not
due to the use of different models, as evidenced by the lower panel of Fig.~\ref{f_newMoehler}, where no systematic is
found. Moreover, the difference is present at any temperature, even for stars cooler than 20\,000~K, where the same models
were used in both works. \citet{Moehler11} used three of our four $\ion{He}{I}$ lines, and we verified that repeating our
measurements with their line list only caused random changes on the derived helium abundance, with an rms of $\sim$0.03~dex.
This is therefore not the cause of the observed systematic, which should be related to the different resolution. Very
interestingly, \citet{Moni09} suggested that their helium abundances could be systematically higher, by a quantity similar
to what was found here, compared to the collection of literature measurements of field stars presented by \citet{OToole08},
which is based on spectra of higher resolution \citep{Edelmann03,Lisker05,Stroer05,Hirsch08}. \citet{Moni09} used the same
instrument and resolution as in the present work. There should therefore be a systematic difference between measurements
based on data with different resolution, with the lower resolution returning higher helium abundances by about 0.2--0.25~dex.
More detailed investigations would be needed to further analyze this behavior.

\subsection{Reddening}
\label{ss_reddening}

\begin{figure}
\begin{center}
\includegraphics[width=9cm]{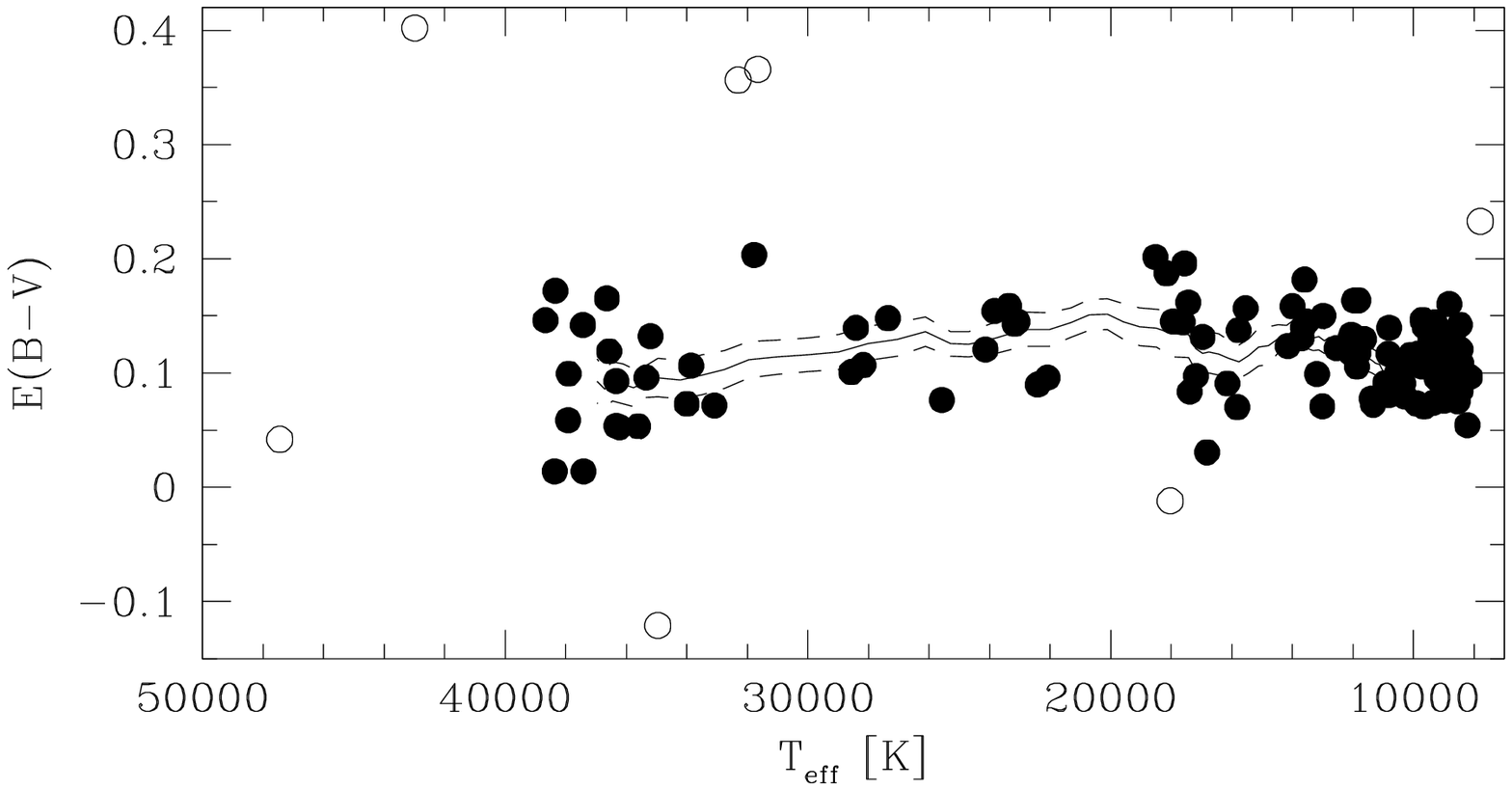}
\includegraphics[width=9cm]{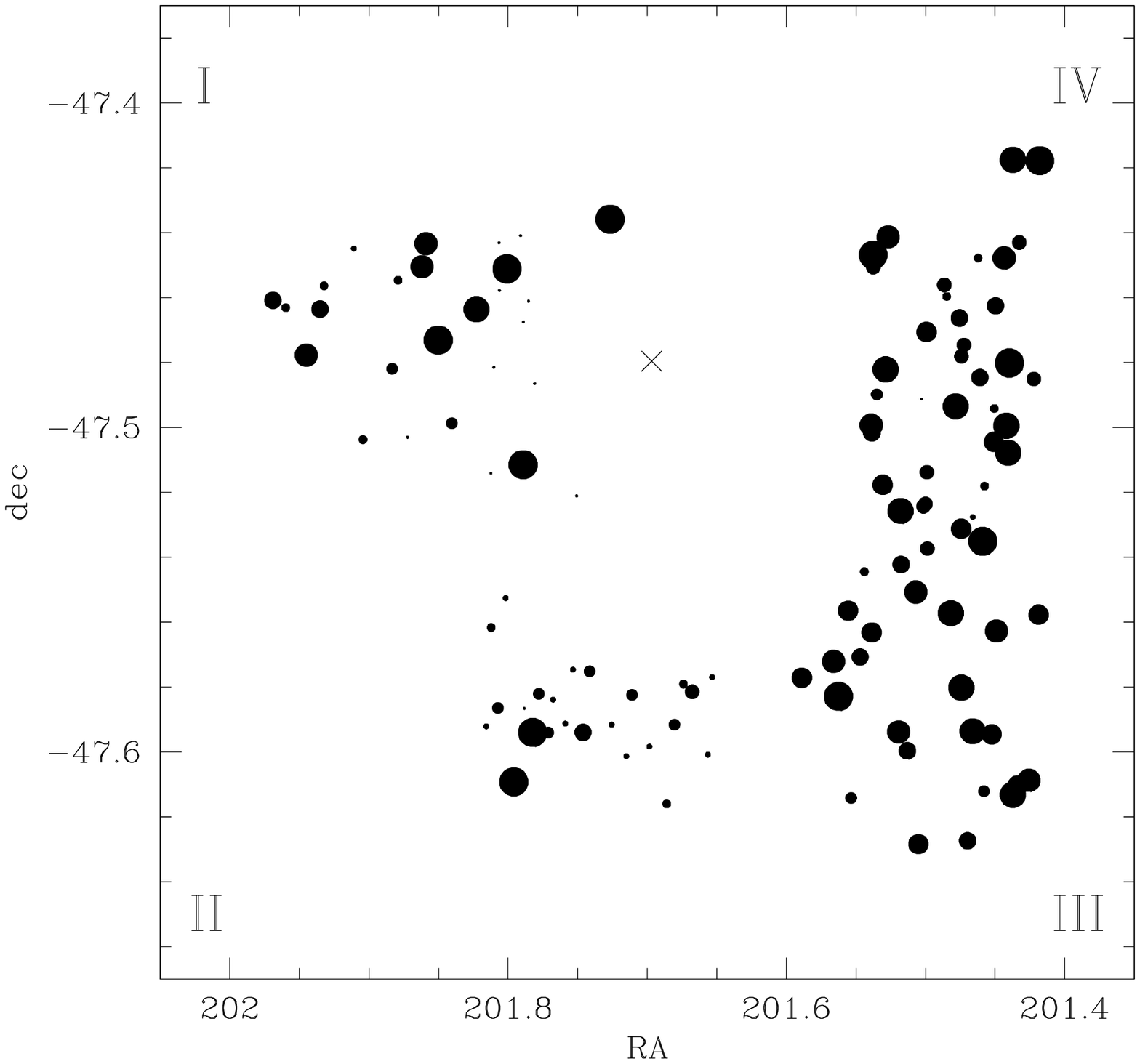}
\caption{Upper panel: reddening estimates for program stars as a function of temperature. The deviant points are
indicated with empty circles. The trend of the mean reddening (calculated as described in the text) with its 1-$\sigma$
stripe is shown with full and dashed lines, respectively. Lower panel: spatial distribution of the observed stars,
where the size of each plotted point is proportional to the star's reddening. The cross indicates the cluster center.}
\label{f_reddening}
\end{center}
\end{figure}

We estimated the reddening of each target, comparing the observed ($B-V$) color from the \citet{Bellini09} catalog
with the theoretical color of a star with the same temperature and gravity. This value was determined interpolating
the \citet{Kurucz93} grid of models with the same metallicity used in the spectra fitting procedure. The results
are shown in the upper panel of Fig.~\ref{f_reddening}, where seven stars deviating from the general trend
are evidenced. The surface parameters of these stars are not peculiar, but we notice that half of them have very low
spectroscopic mass (0.2--0.3~$M_{\sun}$, see Fig.~\ref{f_mass}) and the three targets with $E(B-V)\geq$0.3 are all
redder than the mean HB in Fig.~\ref{f_cmd} ($V\geq$18, $U-V\geq -1$). After their exclusion, the mean reddening is
$E(B-V$)=0.115$\pm$0.004, and the stars are scattered about this value with an rms of 0.037 magnitudes. These values
excellently agree with previous investigations, which found $E(B-V$)=0.11--0.12
\citep{Lub02,Bedin04,Calamida05,Cassisi09}. In Fig.~\ref{f_reddening}, we also show the trend of the mean reddening,
which was calculated by substituting to each target the mean value of the ten adjacent stars in temperature order.
The 1-$\sigma$ stripe was calculated from the statistical error-on-the-mean of this estimate. Moreover, $E(B-V$) shows
only a tiny trend with temperature, which is not significative within errors. The reddening effect is expected to
increase for bluer spectral types \citep{Grebel95}, and in fact the mean value increases from 0.110$\pm$0.005 for
stars between 8\,000 and 10\,000~K to 0.123$\pm$0.008 in the range 20\,000--30\,000~K. Among stars hotter than
32\,000~K, we find a mean reddening that is slightly lower (0.098$\pm$0.012), but still consistent within errors with
the cooler stars. In any case,
the temperature offset discussed in Sect.~\ref{ss_Moehler} cannot be related to this, because a temperature
overestimate would cause a higher reddening. The great majority of these stars are helium rich (86\%, see
Sect.~\ref{ss_hehot}), and it could be argued that the derived lower reddening is due to this discontinuity in
the surface chemical composition. However, we found no evidence that $E(B-V$) is on average higher for the three
helium-poor objects in this temperature regime. On the other hand, it must be taken into account that the theoretical
($B-V$) color has been obtained from a grid of LTE models that should be inadequate for hotter stars. This could have
caused the small offset observed for these targets.

In conclusion, the measured reddening agrees with the literature, confirming that the spectroscopic temperatures
are good. For example, a temperature scale hotter by 10\% would have caused an overestimate of reddening by
0.04~magnitudes at 10\,000~K. We therefore do not find the problems that \citet{Moni07} presented in NGC\,6752.
Hence, we exclude the theoretical colors of the ATLAS9 grid as the origin of their reddening underestimate.

The derived reddening estimates can be used to analyze the pattern of $E(B-V$) in the cluster area. The position
of the target stars, with symbols proportional to the measured reddening, is shown in the lower panel of
Fig.~\ref{f_reddening}. The resulting map is far from uniform, because the stars in the eastern side of the
cluster are on average less reddened than in the western half. The average value of $E(B-V$) is 0.127$\pm$0.05
and 0.133$\pm$0.04 in the fourth and third quadrant, respectively, while it is 0.11$\pm$0.02 in the first and
0.097$\pm$0.008 in the second one. The reddening is therefore uniform in the western half of the cluster, within
the limits of our accuracy. On the eastern side, on the contrary, the reddening is lower, in particular, in
the southeast quadrant. This observed pattern is not a consequence of an uneven distribution of the hotter stars,
whose measured reddening is slightly lower than the average (see the upper panel of Fig.~\ref{f_reddening}, and
discussion above in this section). In fact, we have an equal number of BH candidates in both the eastern and
western half, and none in the second quadrant, where the average reddening is the lowest. A more detailed map of
the cluster differential reddening is prevented by the limited number of data points and by the non-uniform
area sampling. Despite our limitations, our results match the pattern found by \citet{Calamida05}, who found
the clumpy structure of reddening in the direction of $\omega$\,Cen, with more reddened stars clustering on the
west side, and in particular in the northwest quadrant.

\subsection{Helium abundance: canonical blue HB stars ($T_\mathrm{eff}\leq$32\,000~K)}
\label{ss_hecool}

\begin{figure}
\begin{center}
\includegraphics[width=9cm]{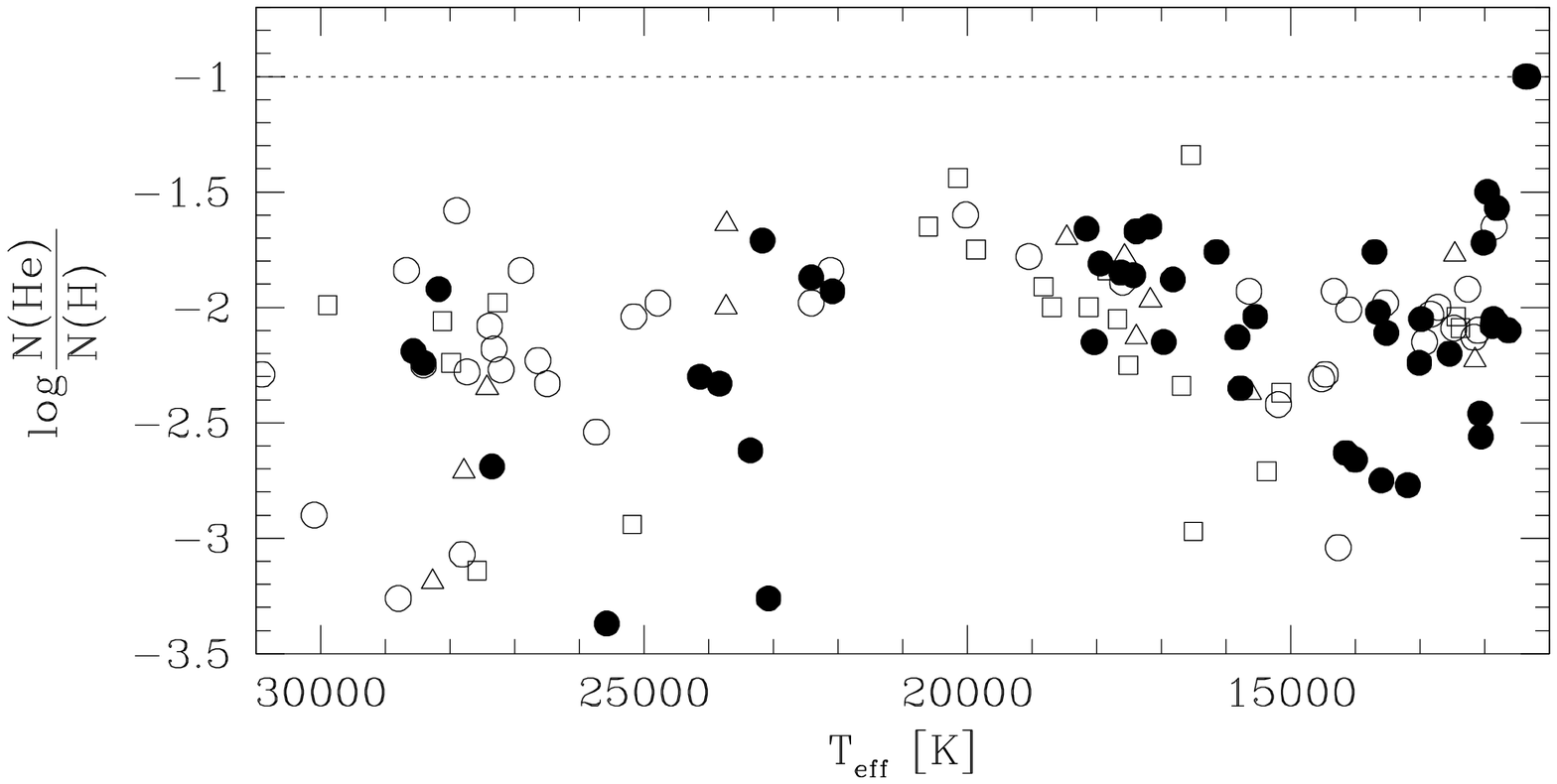}
\includegraphics[width=9cm]{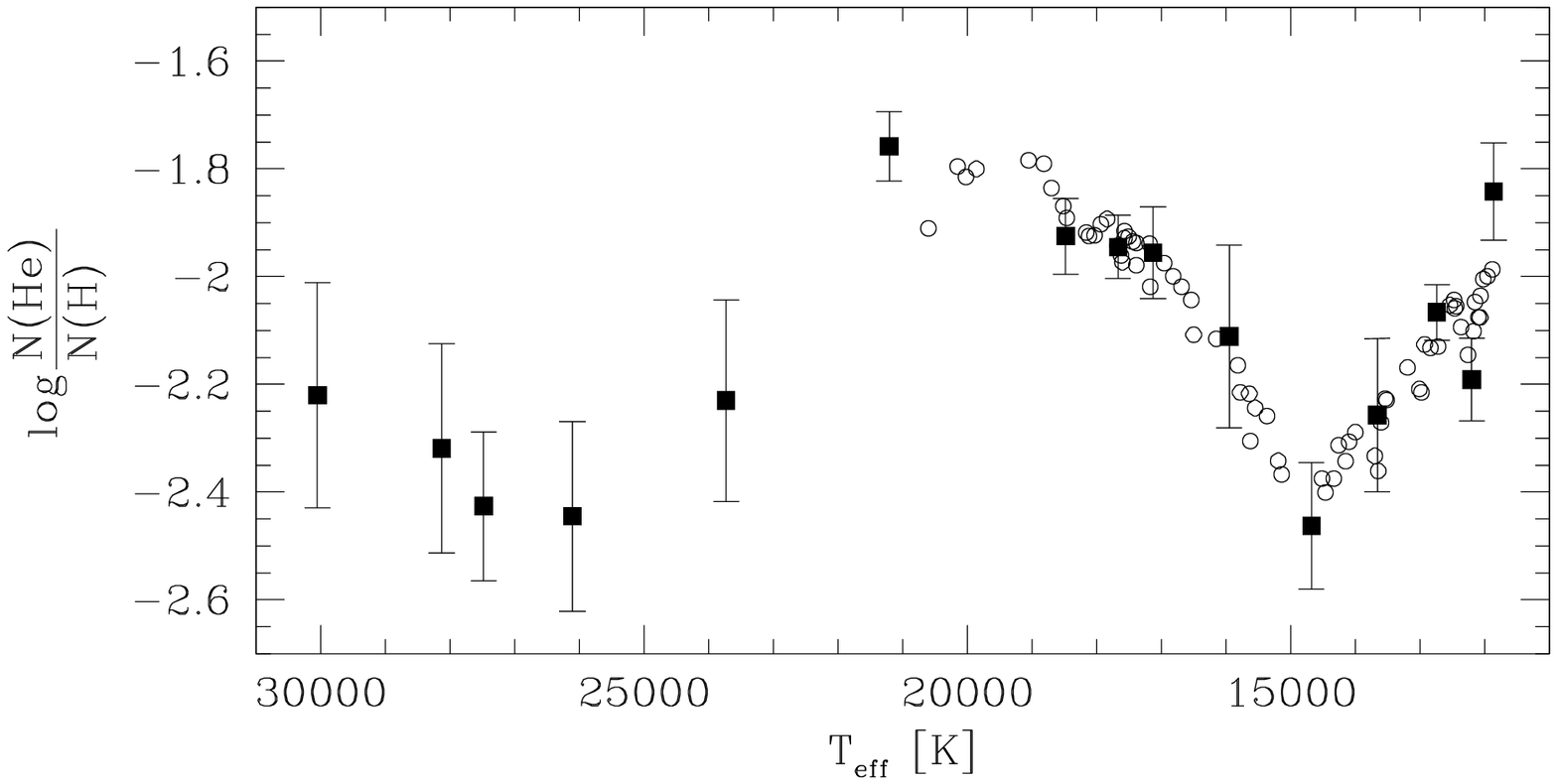}
\caption{Upper panel: surface helium abundance of cooler stars as a function of temperature. The dotted line indicates
the solar value ($Y$=0.25). Filled circles: this work;
empty circles: NGC\,6752 \citep{Moni07}; empty triangles and squares: NGC\,5986 and M\,80, respectively \citep{Moni09}.
Lower panel: trend of surface helium with temperature, obtained binning the data of the upper panel with two
different binning schemes (see text for details).}
\label{f_hecool}
\end{center}
\end{figure}

The surface helium abundance of target stars cooler than 32\,000~K is plotted in the upper panel of
Fig.~\ref{f_hecool} as a function of the effective temperature. The results of \citet{Moni07} and \citet{Moni09}
in NGC\,6752, M\,80, and NGC\,5986 are also plotted, while we exclude the measurement of \citet{Moehler11} from the
comparison, because of the offset discussed in Sect.~\ref{ss_Moehler}. HB stars in $\omega$\,Cen follow the same
trend as their counterparts in the other clusters, despite the peculiarities described in \citetalias{Moni11a}. As
well known, the atmospheric diffusion erases the chemical differences between stars in clusters of different
metallicity \citep{Behr03,Pace06}.

\begin{figure}
\begin{center}
\includegraphics[width=9cm]{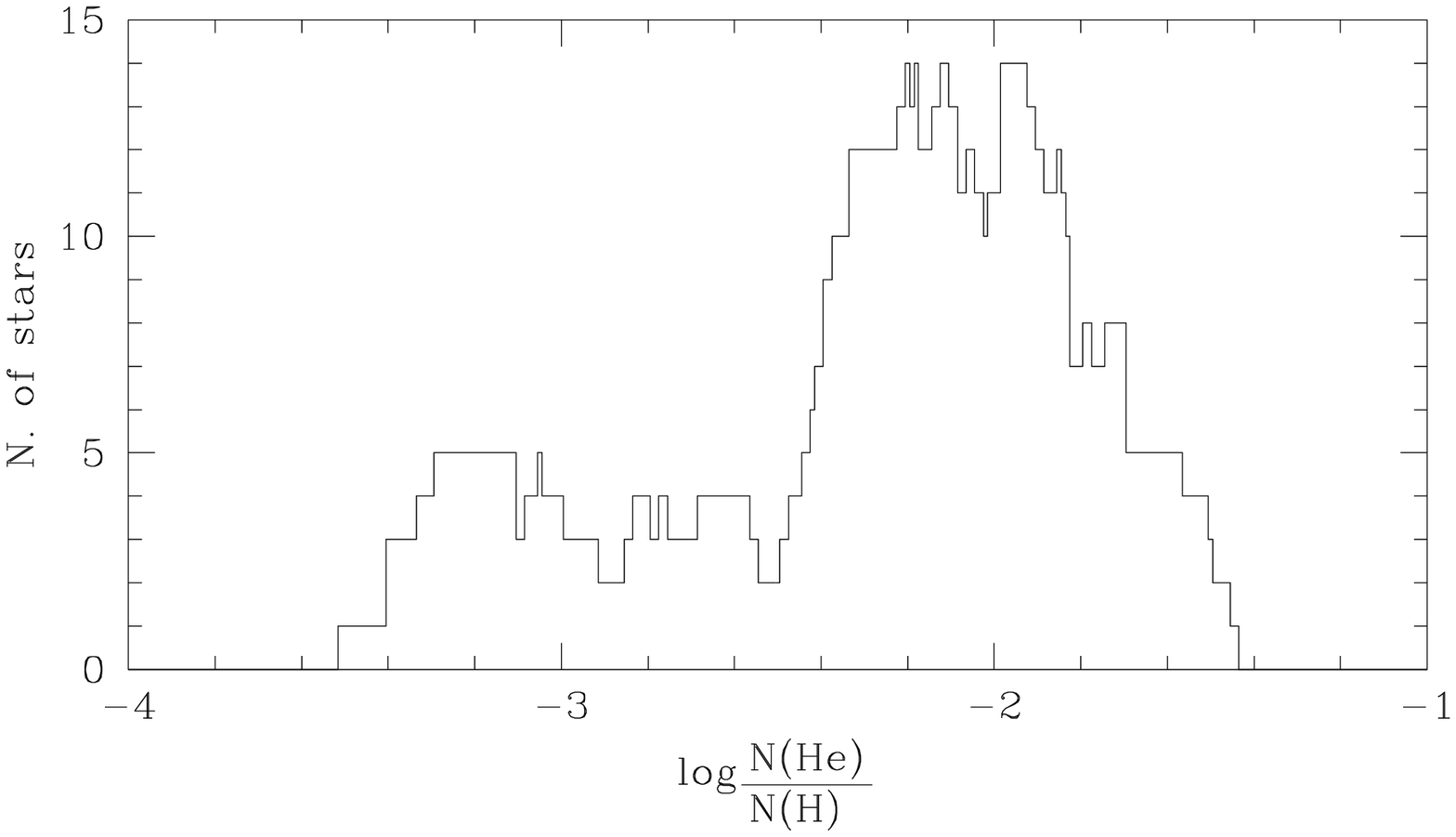}
\caption{Histogram of the distribution of the surface helium abundance of EHB stars ($T_\mathrm{eff}\geq 20\,000~K$)
in $\omega$\,Cen,  NGC\,6752 \citep{Moni07}, NGC\,5986, and M\,80 \citep{Moni09}.}
\label{f_histohe}
\end{center}
\end{figure}

The surface helium abundance of the targets between 11\,500 and 20\,000~K follows the trend discovered by
\citet{Moni09}. This is seen clearly in the lower panel of Fig.~\ref{f_hecool}, where the measurements in all the
clusters are averaged following two binning schemes. In the first scheme, shown with full squares, the targets are binned
in non-overlapping groups of 15 stars each. The error bars indicate the statistical error-on-the-mean in each bin. The
empty circles, on the contrary, show the result of substituting each data point with the average of the ten adjacent
points, in order of temperature. As known for many decades \citep{Baschek75,Heber87,Glaspey89}, the atmosphere of HB
stars hotter than 11\,500~K is depleted in helium because it settles toward deeper layers as an effect of diffusion
\citep{Greenstein67}. The observed change is, however, not abrupt, because the surface He abundance smoothly decreases
with temperature up to $T_\mathrm{eff}\approx$15\,000~K, where it reaches a minimum. Beyond this temperature, the
helium abundance reverts its trend, slowly increasing again up to $\approx$20\,000~K. This behavior can be observed
even in \citet[][see their Fig.~5]{Moehler03} and partially in \citet{Behr03} and \citet{Fabbian05}. A similar
turnoff of the surface iron abundance at $\approx$15\,000~K can be observed in \citet{Pace06}. Hence, the trend is
not a systematic error, but it indicates that the efficiency of the atmospheric diffusion varies with temperature along
the HB, as discussed in detail by \citet{Moni09}.

The results for stars hotter than 20\,000~K are scattered in a wide range of nearly two orders of magnitudes,
much larger than the observational errors (0.1--0.2~dex at these temperatures). The binned data (lower panel of
Fig.~\ref{f_hecool}) suggest a mild increase of the mean abundance for $T_\mathrm{eff} \geq$27\,000~K. This trend is
not significant compared to the dispersion of the points, as indicated by the large error bars associated to the average
values, but it coincides with what has been observed among field stars \citep[see Fig.~1 of][]{OToole08}.
\citet{Edelmann03} discovered a family of extremely helium-poor field EHB stars, comprising $\sim$10\% of the whole
population, with a surface helium abundance 1--1.5~dex lower than the other EHBs. Thus, a bimodal distribution
of $\log{(N(\mathrm{He})/N(\mathrm{H}))}$ should be expected, even in GCs.  The histogram shown in Fig.~\ref{f_histohe}
suggests that this could be the case. It was obtained associating to each value of $\log{(N(\mathrm{He})/N(\mathrm{H}))}$
the quantity of stars with helium abundance within $\pm$0.15~dex, which is the mean observational error at these
temperatures. The distribution is relatively smooth, but at least two main peaks are visible, at about
$\log{(N(\mathrm{He})/N(\mathrm{H}))}=-2$ and $-$3.2~dex, A gap, or even a third peak, could be present at about
$-$2.8~dex. Two out of the fourteen EHB stars in $\omega$\,Cen have $\log{(N(\mathrm{He})/N(\mathrm{H}))}\leq-2.8$,
corresponding to 14\%, compared to 15\% found in NGC\,6752.
These objects are apparently more frequent in M\,80 and NGC\,5986 (22\% and 40\%, respectively), but the statistics
is not significative due to the small observed samples. In conclusion, about 15\% of the EHB stars in the four GCs
have very low surface helium abundance. However, they are not clearly separated from the other more numerous EHB
objects, and the evidence for a bimodal distribution is not conclusive. The position of the two extremely He-poor
EHB stars in $\omega$\,Cen in the CMD and in the temperature-gravity plane is not distinct from the distribution of
the other stars, as shown in Fig.~\ref{f_cmd} and \ref{f_hehottg}.

\subsection{Helium abundance: blue hook candidates}
\label{ss_hehot}

\begin{figure}
\begin{center}
\includegraphics[width=9cm]{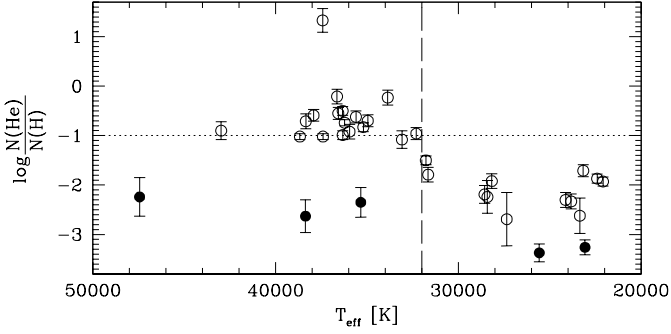}
\caption{Surface helium abundance of the hottest targets as a function of temperature. The helium-depleted stars
discussed in the text are indicated with filled circles, while all the other stars are shown with empty circles.
The dashed line at $T_\mathrm{eff}$=32\,000~K separates the EHB stars from the BH candidates, while the dotted line
indicates the solar helium abundance ($Y$=0.25).}
\label{f_hehot}
\end{center}
\end{figure}

The surface helium abundance of our hottest targets is shown in Fig.~\ref{f_hehot}. Two distinct groups of stars are
clearly visible: the atmosphere of few stars with $\log{(N(\mathrm{He})/N(\mathrm{H}))}\leq-2.2$ is depleted in helium
by more than one dex with respect to the majority of the stars, which exhibit a solar or super-solar abundance
($\log{(N(\mathrm{He})/N(\mathrm{H}))}\geq-1.1$). While \citet{Moehler11} found that 28\% of their stars in the range
30\,000-50\,000~K is He-poor, we detect a lower fraction (14\%). This difference could partially be an effect of a
small number statistics, but it is moderately significative because, given the statistics of \citet{Moehler11}, our
results have a probability of 10\% of being due to pure chance. Moreover, we notice that \citet{Moehler11} measured
even a fraction of extremely He-poor EHB stars ($\log{(N(\mathrm{He})/N(\mathrm{H}))}\leq-3$,
$T_\mathrm{eff}$=20\,000-32\,000~K) a factor of two higher ($\sim$40\%) than in the present work. The discrepancy
could be reduced if, strictly adopting the temperature range used by \citet{Moehler11} and taking into account the
offset of $\sim$0.2~dex in He abundance between the two works discussed in Sect.~\ref{ss_Moehler}, we
considered the two stars at $\sim$31\,000~K and $\log{(N(\mathrm{He})/N(\mathrm{H}))}\approx-1.7$ as He-depleted BH
objects. However, comparing Fig.~\ref{f_hehot} with the trend of field stars \citep[Fig.~1 of][]{OToole08}, these
targets appear most likely as the connection between the EHB and BH sequences richer in helium. This kind of
transitional objects is missing in the \citet{Moehler11} sample.

\begin{figure}
\begin{center}
\includegraphics[width=9cm]{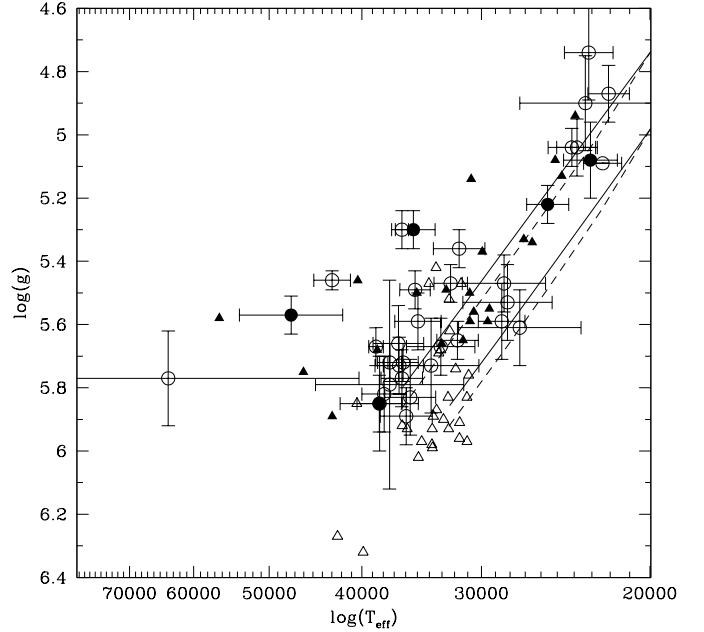}
\caption{Position of the hottest targets in the temperature-gravity plane. Big circles: this work; small triangles: data
from \citet{Moehler11}. Empty and full symbols refer to the helium abundances of the objects as defined in the text
and shown in Fig.~\ref{f_hehot}. The full and dashed lines show the position of the canonical zero-age and terminal-age
HB for canonical and helium-enriched stars, respectively.}
\label{f_hehottg}
\end{center}
\end{figure}

\citet{Moehler11} showed that the helium-poor hot stars in their sample are most likely evolving off the HB. In fact,
the evolutionary path of post-EHB stars, after the exhaustion of helium in the core, draws the stars toward lower
gravities and higher temperatures \citep[see, for example,][]{Moehler04}, while they become bluer and brighter. The
position of two of our helium-poor very hot targets in the CMD and in the temperature-gravity plane, shown in
Fig.~\ref{f_cmd} and~\ref{f_hehottg} respectively, differs from the bulk of BH objects, indicating that they also are
most probably evolving toward the white dwarf cooling sequence. A third helium-depleted object (\#181428) is, on the
contrary, not distinct to the helium-rich stars, and its nature is more uncertain. As already discussed in
Sect.~\ref{ss_Moehler}, Figure~\ref{f_hehottg} suggests that an offset between our measurements and those of
\citet{Moehler11} at the hottest end of the distribution could be present, because their helium-rich targets at
$\log{g}\geq5.8$ cluster on the theoretical zero-age HB (ZAHB), while ours do not.

\subsection{Spatial distribution of BH stars}
\label{ss_spatial}

\begin{figure}
\begin{center}
\includegraphics[width=9.5cm]{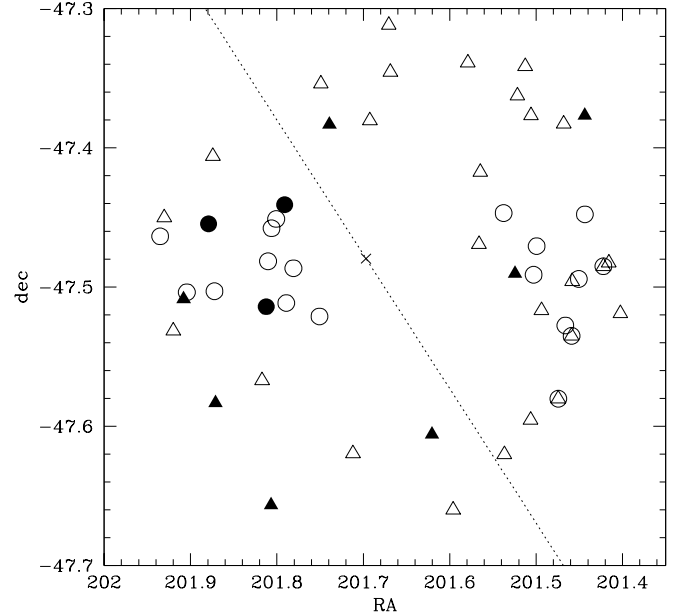}
\caption{Spatial distribution of the BH stars. The symbols are as in Fig.~\ref{f_hehottg}. The cross shows the position
of the cluster center, while the dotted line cuts the cluster area as defined in the text.}
\label{f_spatial}
\end{center}
\end{figure}

\citet{Moehler11} found an asymmetric spatial distribution for their helium-rich BH stars, which are more concentrated
in the northwestern part of the cluster. Their helium-depleted targets, on the contrary, do not show such asymmetry.
The spatial distribution of our targets and those from \citet{Moehler11} are shown in Fig.~\ref{f_spatial}. In both
cases, however, the cluster is not uniformly sampled, and this introduces selection effects that must be taken into
account. In fact, \citet{Moehler11} also found that the helium-rich stars in the northwestern half
of the cluster outnumber those in the other half by a factor of 2.7, but in that area they observed twice the quantity of
targets than in southeastern half. However, the relative quantity of helium-poor to helium-rich stars should not be
affected by any bias, because the photometric selection criteria are uniform in the whole cluster area. The helium-poor
stars found by \citet{Moehler11} are more frequent in the southeast side of the cluster (42\%) with respect to the other
half (14\%).

Our sample consists of two distinct groups of stars of approximatively the same quantity of stars, one east and the
other west of the cluster center. As the helium-rich stars dominate the sample (Sect.~\ref{ss_hehot}), it is no
surprise that we find a similar quantity on both sides of the center (nine stars, both in the east and west group), and
no asymmetry is detected. However, the three helium-poor objects are found on the east side only. Dividing the cluster
along a line with position angle 35$\degr$ counted from north toward east, we thus detect a high fraction of helium-poor
stars in the southeast half (25\%), while no such object is detected out of nine stars in the other half. Hence, if we
analyze their frequency instead of their absolute quantity, our results confirm those of \citet{Moehler11}. Merging the
two samples and taking into account that four targets are in common, the resulting fractions are 11\% (three out of
28 stars) and 30\% (seven out of 23), in the northwest and southeast half, respectively. However,
the evidence for an asymmetric distribution of the helium-poor targets is only mildly significative: a two-dimensional
Kolmogorov-Smirnov test, performed following the recipes of \citet{Peacock83}, reveals that there is a $\sim$11\%
probability that they are randomly drawn from the distribution of the observed sample. We conclude that the present
observations suggest that an intriguing asymmetry could be present, but the result is not conclusive. A more systematic
investigation, uniformly sampling the cluster area with a larger sample, would be required to clarify the issue. As
already noted by \citet{Moehler11}, the suspected asymmetry is similar to the pattern of differential reddening
evidenced by \citet{Calamida05}, discussed in Sect.~\ref{ss_reddening}, with helium-poor stars being more frequent in
the less reddened part of the cluster. This connection is not straightforwardly explained. A selection effect could be
introduced by the photometric target selection, if helium-poor stars are on average bluer than the helium-rich objects,
but still the differential reddening that we find ($\Delta E(B-V)\approx$0.03-0.04) is not large enough to hide blue
objects in the more reddened regions.

\subsection{Masses}
\label{ss_mass}

\begin{figure}
\begin{center}
\includegraphics[width=9.5cm]{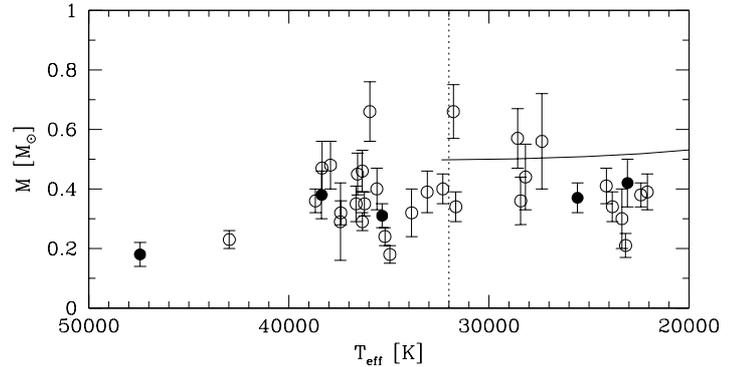}
\caption{Spectroscopic masses of the hottest targets as a function of temperature. The full line shows the canonical
model expectations. Empty and full symbols are used for helium-rich and helium-poor stars, respectively, as defined
in the text and shown in Fig.~\ref{f_hehot} with the same symbols.}
\label{f_mass}
\end{center}
\end{figure}

In \citetalias{Moni11a}, we discussed the problem of the strong underestimate of the masses calculated for the stars cooler
than 32\,000~K. Analogous results were also obtained by \citet{Moehler11}, and the interpretation of this systematic
was not straightforward. In Fig.~\ref{f_mass} we show the results for the hotter end of our sample. Our mass estimates
are too low even for the hot targets, contrary to \citet{Moehler11}, whose measurements agreed with the model expectations.
Our underestimate could be easily explained by the offset in temperature and/or surface gravity for our hottest stars,
discussed in Sect.~\ref{ss_Moehler} and suggested by Fig.~\ref{f_hehottg}. According to Eq.~(\ref{eq_mass1}), either
a temperature overestimate by $\sim$3\,000~K or a gravity underestimate by $\sim$0.14~dex causes an underestimate of the mass
by about 35\%. Correcting an offset of this magnitude would increase the average mass of these stars from $\sim$0.37 to
$\sim 0.5~M_{\sun}$, in agreement with \citet{Moehler11} and with the theoretical expectations. However, the contemporary
presence of both systematics, as suggested by Fig.~\ref{f_Moehler}, would result in a mass overestimate for the hottest
targets of $\sim 0.23~M_{\sun}$ on average.

\citet{Moni07} discovered eight EHB stars in NGC\,6752 with anomalously high spectroscopic mass, which possibly occupy
redder and fainter loci in the CMD. Identical results were also found in M\,80 \citep{Moni09}. In $\omega$\,Cen, we
detect three EHB stars out of 14 (21\%) with spectroscopic mass more than 0.1~$M_{\sun}$ higher than the other
targets at the same temperature. All these objects have a surface helium abundance higher than
$\log{(N(\mathrm{He})/N(\mathrm{H}))}= -2.7$, and none belongs to the extremely helium-poor sub-population. Their position
in the cluster CMD is shown as filled triangles in Figure ~\ref{f_cmd}. They are, indeed, on average, fainter and/or redder
than the other main HB population. We detect a lower fraction of anomalous stars with respect to NGC\,6752 and M\,80,
where they constituted the $\sim$40\% of the analyzed sample in both cases. However, this difference is not very
significative, because the probability of detecting only three anomalous stars in our sample, assuming that they are the
40\% of the population, is 12\%.

\subsection{Color-temperature relation}
\label{ss_colortemp}

\begin{figure}
\begin{center}
\includegraphics[width=9.5cm]{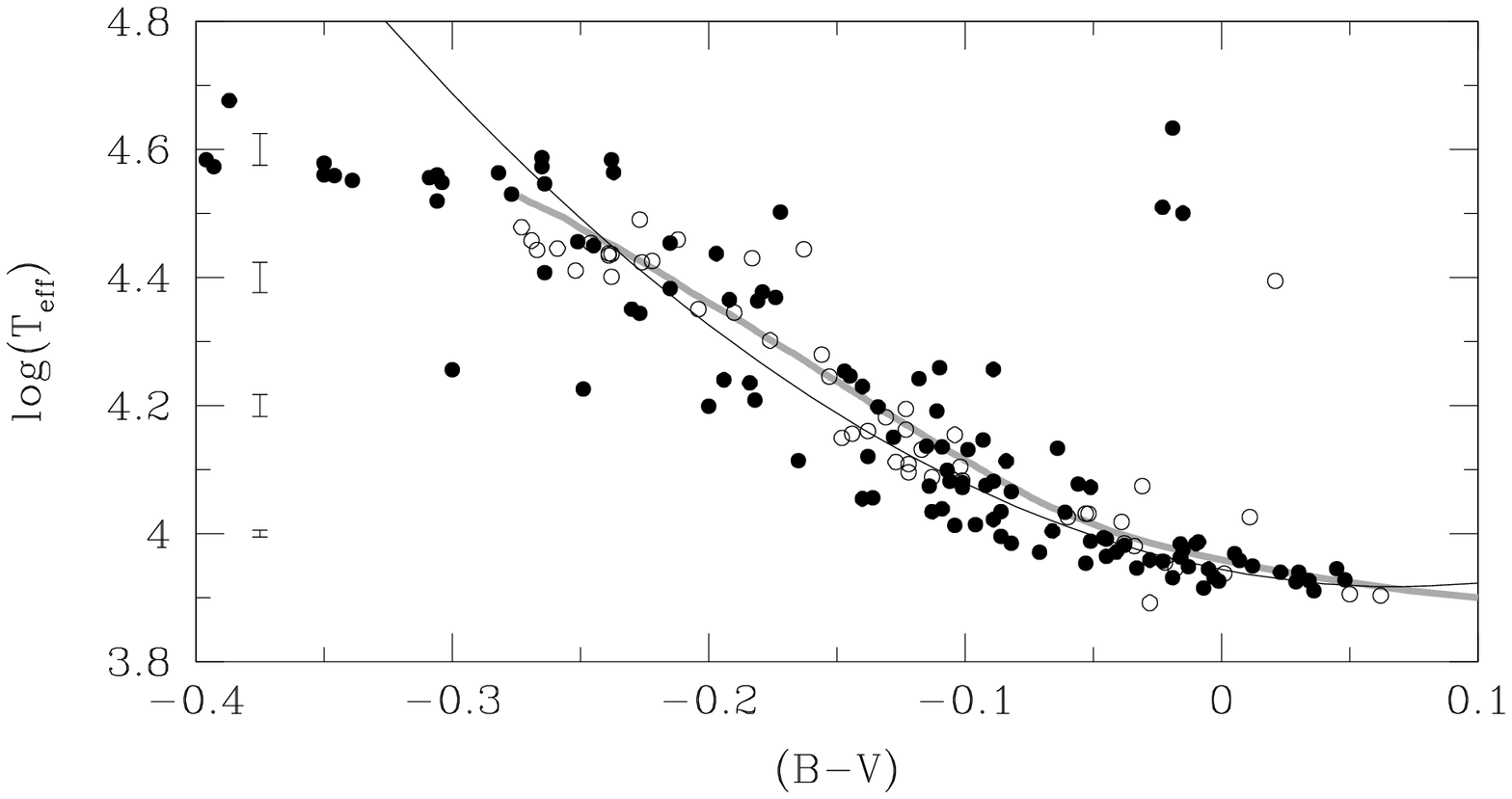}
\includegraphics[width=9.5cm]{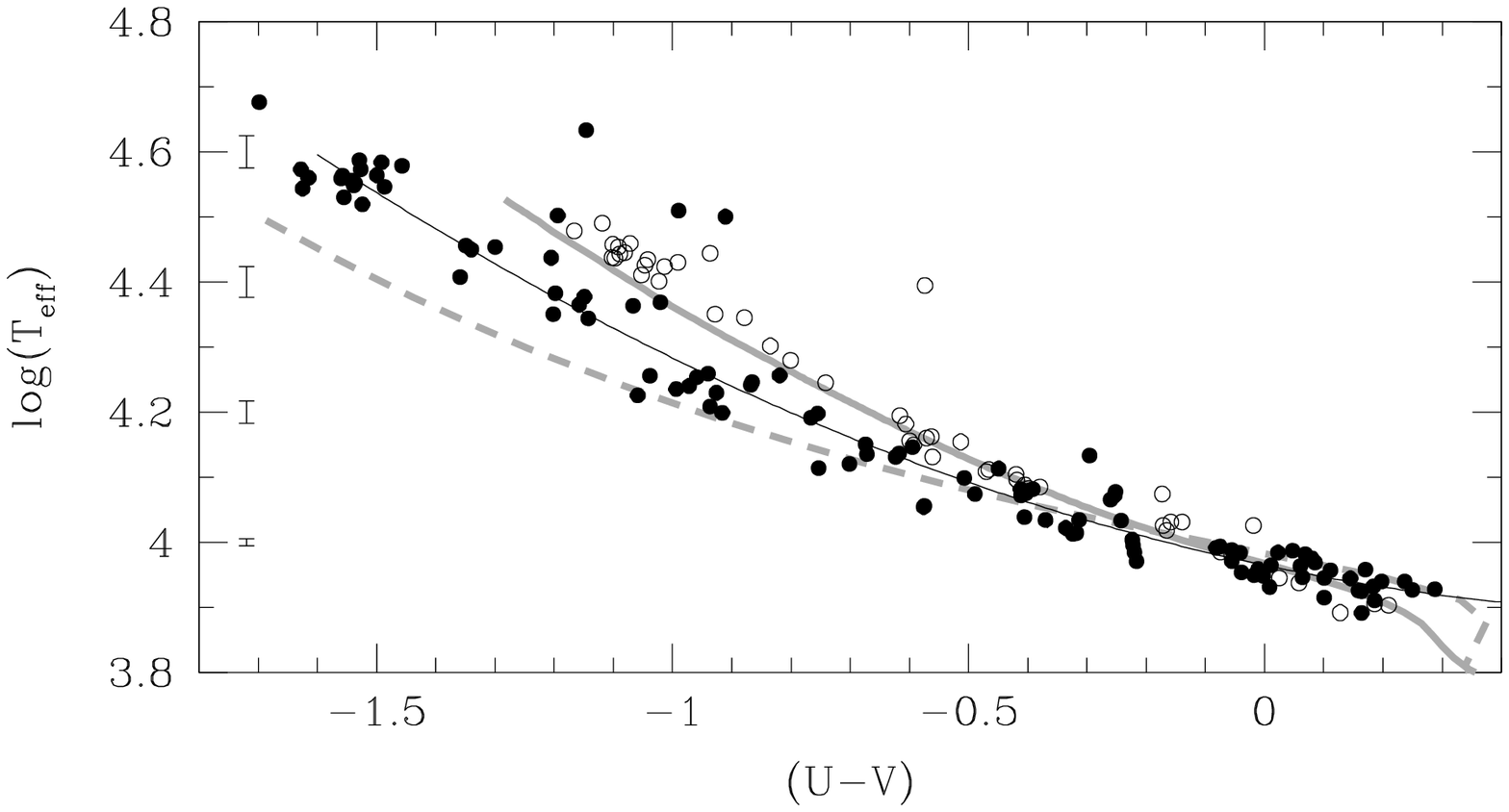}
\caption{Photometric color of the program stars as function of their measured temperature (filled circles). The black
curve indicates the fit of the data, Eq.~(\ref{eq_TeffUV} and (\ref{eq_TeffBV}). The grey thick curve shows the
theoretical zero-age HB of \citet{Cassisi09}. The same model, re-derived for the non-standard filters used in the
adopted photometry, is shown with a dashed grey curve. Empty circles show the position of NGC\,6752 stars \citep{Moni07}.}
\label{f_colortemp}
\end{center}
\end{figure}

In Fig.~\ref{f_colortemp}, the spectroscopic temperature of the target stars is plotted against their observed color.
This was dereddened assuming $E(B-V$)=0.115, as measured in Sect.~\ref{ss_reddening}, and $E(U-V$)=0.194 from the
\citet{Cardelli89} transformations of reddening in different bands. After the exclusion of some deviating stars, the
fit of the data points in Fig.~\ref{f_colortemp} returns the analytical expressions
\begin{equation}
\log{(T_\mathrm{eff})}=3.9646-0.1920\cdot (U-V)+0.1265\cdot (U-V)^2,
\label{eq_TeffUV}
\end{equation}
\begin{equation}
\log{(T_\mathrm{eff})}=3.9438-0.7769\cdot (B-V)+5.6725\cdot (B-V)^2.
\label{eq_TeffBV}
\end{equation}
The rms of the residuals in $\log{(T_\mathrm{eff})}$ is 0.035~dex for Eq.~\ref{eq_TeffUV}, and 0.070 for
Eq.~\ref{eq_TeffBV}. The ($B-V$) color is, as expected, much less sensitive to temperature than ($U-V$), as
indicated by the steeper solution.

In  Fig.~\ref{f_colortemp}, the  theoretical ZAHB from \citet{Pietrinferni06} and
\citet{Cassisi09} with $Z$=0.001 and $Y$=0.246 for the $\alpha-$enhanced mixture is overplotted to the data as a grey
solid curve. Fifty-one HB stars of NGC\,6752 are also plotted,
whose spectroscopic temperatures and colors are taken from \citet{Moni07} and \citet{Momany02}, respectively.
\citet{Moni07} measured a too low reddening for these stars, arguing that the photometric data could suffer from a
zero-point offset. Following their results, ($B-V$) was not corrected for reddening, while the data in the
temperature-($U-V$) plot were shifted horizontally to force the cooler stars ($\log{T_\mathrm{eff}}\leq$4) to
superimpose to the theoretical ZAHB. In any case, what matters here is the trend of temperature with color, which is
not altered by this zero-point correction.

The temperature-($U-V$) relation of $\omega$\,Cen stars cannot be directly compared to the standard theoretical ZAHB,
shown as a solid grey curve. This is because the $U$ filter used by \citet{Bellini09} differs substantially from the
standard $U$ band in the Johnson system \citep[see discussion and Fig.~3 in]{Momany03}. Hence, we re-computed 
the color - ${T_\mathrm{eff}}$ transformations by adopting the transmission curve of
the WFI@2.2m non-standard $U$ filter, following the same approach used in \citet{Momany03} and applied these
transformations to the stellar models by \citet{Pietrinferni06}. The ZAHB sequence transferred from the theoretical HR
diagram to the observational one by using the appropriate set of transformations is shown as a grey dashed line in the
central panel of Fig.~\ref{f_colortemp}.

The observed color-temperature distribution of target stars in NGC\,6752 follows closely the theoretical ZAHB model,
both in the upper and lower panels of Fig.~\ref{f_colortemp}.
A similar agreement is observed for stars in M\,80 and NGC\,5986, using the spectroscopic temperatures of
\citet{Moni09} and the colors of \citet{Momany03} and \citet{Momany04}, after correction of a photometric zero-point
offset. The same trend is found for $\omega$\,Cen stars, when the ($B-V$) color is involved. However, the
agreement between the ZAHB sequence transferred in the observational photometric plane with the appropriate
transformations and observed ($U-V$)-temperature relation is limited only to $T_\mathrm{eff}\leq$12\,000~K in this
cluster. For hotter stars, the temperatures predicted by the model are lower than the ones we derive, and the
difference increases monotonically, reaching up to $\sim$8\,000~K at $T_\mathrm{eff}\approx$32\,000~K.

The trend of reddening with temperature (discussed in Sect.~\ref{ss_reddening}) is too small to account for the
disagreement between observed data and the re-calculated ZAHB model. Similarly, the assumed $E(U-V$) cannot be the
cause of the disagreement, since it can only shift horizontally the distribution of the data points. 
We also verified whether a reddening correction depending on the stellar $T_\mathrm{eff}$ value could, at least
partially, account for the mismatch between theory and observations. For this aim, we adopted the approach
outlined in \citet{Bedin09}. However, due to the not large (average) reddening of $\omega$ Cen, accounting for a reddening
correction that depends on the $T_\mathrm{eff}$  would affect the theoretical prediction on the $(U-V)$ color at the level
of a negligible $\sim0.02$ mag.

Not even the temperature offset discussed in Sect.~\ref{ss_Moehler} can account for the disagreement between the
observed and theoretical temperature-($U-V$) relation. Indeed, the use of different model spectra introduces only a
too small offset ($\sim$400~K), which anyway shows a constant trend with temperature. On the other hand, the mismatch
between the temperatures of our hottest targets and those derived by \cite{Moehler11} is of the same order of their
offset with respect to the theoretical temperature-color relation. However, Moehler et al.'s temperature estimates
excellently agree with ours for stars cooler than 20\,000~K, where the offset in Fig.~\ref{f_colortemp} is clear.
Moreover, a temperature offset would also affect the $T_\mathrm{eff}$-($B-V$) relation, contrary to what was observed.
It might also affect NGC\,6752 stars, because \citet{Moni07} employed the same models used in this study.

It is quite natural to attribute the mismatch to the onset of atmospheric diffusion at 11\,000--12\,000~K
\citep{Grundahl99}, which alters the surface abundance of HB stars. Nevertheless, this interpretation presents a
series of problems. First, the difference between the measured temperatures and the theoretical model does not
follow the trend of the diffusion discussed in Sect.~\ref{ss_hecool}, which reaches a maximum at about 15\,000~K
and then fades out at higher temperatures. In addition, we checked the impact of diffusive
processes on the location of the theoretical ZAHB locus in the $T_\mathrm{eff}-(U-V)$ plane. This was done by
following the approach used in \citet{Dalessandro11}: we mimic the effect of diffusive processes by applying
bolometric corrections appropriate for solar iron abundance when $T_\mathrm{eff}$ is larger than 12,000 K. However,
we find that, at least for the $(U-V)$ color, the impact of diffusive processes on the predicted $T_\mathrm{eff}$
-color relation is quite negligible, i.e., the ZAHB loci transferred in the observational plane by accounting (or not)
for the occurrence of diffusive processes overlap almost perfectly. Finally, diffusion and radiative levitation are
obviously also active in HB stars of NGC\,6752 \citep{Moehler00,Moni07}, but they behave very differently.

One cannot exclude that the mismatch is simply due to a systematic introduced by the use of a non-standard
$U$-filter. If this is the case, then even the presumably simple photometric calibration process might hide some
unexpected results. In \citet{Momany03} it was shown that the use of a slightly bluer $U$-filter (with respect to
the standard one, see their Fig.~3) triggers the onset of an unphysical phenomenon: HB stars bluer
than the instability strip can display ($U-B$) colors redder than that of red giants at the same luminosity level.
Indeed, the non-standard $U$-filter employed in the $\omega$\,Cen data set, reduced first by \citep{Momany03}
and later by \citet{Bellini09}, covers a bluer wavelength range and misses almost completely the Balmer jump.
This was shown to imply a fainter $U$-magnitude for the blue HB stars, which translated to redder ($U-B$) colors
(the so-called ``BHB  red incursion'' \citealt{Momany03} or ``ultraviolet deficiency'' as in \citealt{Markov01}).
We verified that the observed $T_\mathrm{eff}-(U-V)$ relation is identical when the \citep{Bellini09}
photometric data are replaced with magnitudes from the \citep{Momany03} catalog, except for a zero-point offset
of about $\approx$0.1~magnitudes, which only causes a horizontal shift of the points in Fig.~\ref{f_colortemp}.
This behavior is not surprising, since the photometric calibration of \citet{Bellini09} was performed by means
of the \citet{Stetson00,Stetson05} secondary standards except for the $U$-filter, which was calibrated using the
\citet{Momany03} catalog, anchored to the \citet{Landolt92} system of standards.

The unavoidable conclusion is that the hottest stars in $\omega$\,Cen are fainter than expectations in the $U$
band. In fact, if a similar effect was also present in another band, or if the temperatures were overestimated,
or if the stars were really hotter than expected, this would have been easily detected in the study of the reddening
in Sect.~\ref{ss_reddening} and in the ($B-V$)-temperature relation. A similar conclusion was reached for the
blue HB at the level of the \citet{Grundahl99} jump, and this feature was successfully correlated to the use of
non-standard $U$-filters. We cannot draw a similar solid conclusion for the hottest $\omega$\,Cen stars nor
totally exclude that this is yet another new peculiarity of the $\omega$\,Cen endless puzzle. In any case, the
connection of this effect with the anomalous results discussed in \citetalias{Moni11a} is not straightforward, even
if a temperature hotter than expectations could indeed explain both the observed offset in the temperature-gravity
plane and the underestimated masses. More investigation is needed to find a comprehensive explanation of
all these observations.

Before closing this section, we wish to note that since $\omega$ Cen hosts a stellar sub-population hugely enhanced
in He abundance (see \citealt{King12} and references therein), one has to check if the disagreement between the
location of the GC hot HB stars with respect the theoretical ZAHB computed by accounting for a canonical He
abundance could be partially accounted for by considering a ZAHB locus corresponding to an He-enhanced stellar
population. We verified this scenario by comparing a ZAHB locus for $Y$=0.40 with the empirical data. However,
at the cooler $T_\mathrm{eff}$ values where the He-enhanced ZAHB is redder than the ZAHB for the canonical He
content, the two ZAHBs overlap almost perfectly in the hot temperature regime. As a consequence, the occurrence of
the multiple population phenomenon in $\omega$ Cen cannot help in solving the observed discrepancy between theory
and observations.


\section{Conclusions}
\label{s_conclusions}

We analyzed the fundamental parameters spectroscopically derived for a sample of more than 100 HB stars in
$\omega$\,Cen. Our results can be summarized as follows:
\begin{itemize}
\item We showed that the surface helium abundances measured on low-resolution spectra are systematically higher by
about 0.2-0.25~dex with respect to measurements at higher resolution. The presence of this offset was already
suspected in previous works.
\item We derived a mean cluster reddening of $E(B-V$)=0.115$\pm$0.004, in good agreement with previous estimates.
However, we confirmed the recent discovery by \citet{Calamida05} that the reddening is not uniform across the cluster
area, with our measurements being on average 0.03-0.04~mag higher in the western half than in the southeast region.
\item The surface helium abundance of $\omega$\,Cen HB stars hotter than 11\,500~K is very similar to that of
analogous stars in other clusters, despite their peculiar gravity discovered by \citet{Moni11a}. Diffusion
processes efficiently erase the differences among the atmospheres of stars with different initial chemical
composition. From the measurements of 121 stars in four clusters, we find a clear trend of surface helium with
temperature, which most probably reflects a dependence of diffusion efficiency with temperature. In fact, the helium
abundance decreases with $T_\mathrm{eff}$, reaching a minimum at $\sim$15\,000~K and then increases again for hotter
stars.
\item The surface helium abundance of EHB stars mildly increases with temperature for
$T_\mathrm{eff}\geq$27\,000~K, analogous to the trend observed among field stars, but the measurements are more
scattered than what would be expected for observational errors alone.  In both $\omega$\,Cen and NGC\,6752, where a
significant sample was analyzed ($\geq14$ targets), $\sim$15\% of the EHB stars have a surface helium abundance
$\sim$1~dex lower than the others. This suggests that two families of EHB stars with different helium abundance are
present even in GCs, analogous to what \citet{Edelmann03} observed among field objects. However, evidence that
our EHB stars follow a bimodal distribution is still not conclusive, because it is blurred by observational errors.
\item Two groups of BH candidate stars are observed in our sample: the majority has a solar or super-solar surface
helium abundance, while a small quantity (14\%) is strongly helium depleted ($\log{(N(\mathrm{He})/N(\mathrm{H}))}\leq -2$).
This result confirms what had already been found by previous investigations, both in $\omega$\,Cen \citep{Moehler11}
and in the field \citep{Edelmann03}. However, we detect a fraction of helium-poor objects lower than what had been
previously measured. This lower fraction cannot be due to a different definition
of the transition between the two groups, or by the aforementioned offset in helium abundance because the gap between
the two groups is very large ($\geq$1~dex). Our results are consistent with the scenario where the helium-depleted BH
candidates are post-HB objects evolving off toward the white dwarf cooling sequence, as proposed by \citet{Moehler11}.
\item We do not detect an asymmetric spatial distribution of helium-rich BH stars, at variance with
\citet{Moehler11}. Nevertheless, the helium-poor objects are more frequent in the southeast half by about a factor
of three, both in our and their sample. This finding is not affected by the non-uniform cluster sampling, which
could be introducing selection effects when the spatial distribution of BH stars is analyzed. However, the result is
only marginally significative, because a Kolmogorov-Smirnov test revealed that it has a 10\% probability of
being due to pure chance only.
\item We find indications that EHB stars with anomalously high spectroscopic mass could be present even in
$\omega$\,Cen after their discovery in NGC\,6752 and M\,80 \citep{Moni07,Moni09}. Their fraction is lower than in
the other clusters, but the difference is not significative. Their photometric behavior resembles that of their
analogs in the other clusters. Unfortunately, only three such stars are detected, too few for a more detailed
analysis of their properties.
\item Our targets follow a ($B-V$)-$T_\mathrm{eff}$ relation that is well reproduced by the theoretical HB models.
The empirical ($U-V$)-$T_\mathrm{eff}$ curve, on the contrary, matches the models only for
$T_\mathrm{eff}\leq$11\,000~K, hotter stars being systematically redder than the theoretical expectations. HB stars
in NGC\,6752 do not deviate from the model curve. This cannot be explained by diffusion processes alone and points
to another peculiarity of $\omega$\,Cen HB stars, in addition to those already discussed by \citet{Moni11a}, with
respect to their analogs in NGC\,6752 and other two comparison clusters. We do not have an explanation for this
behavior.
\end{itemize}


\begin{acknowledgements}
C.M.B. acknowledges support by the FONDAP Center for Astrophysics 15010003, BASAL Center for Astrophysics and
Associated Technologies (CATA) PFB-06/2007. SC is grateful for financial support from PRIN-INAF 2009  ``Formation and
Early Evolution of Massive Star Clusters" (PI: R. Gratton) and PRIN-INAF 2011 ``Multiple populations in Globular
Clusters: their role in the Galaxy assembly" (PI: E. Carretta) 
\end{acknowledgements}


\bibliographystyle{aa}
\bibliography{OmegaII.bib}

\longtab{2}{
\begin{longtable}{lcrrccccccc}
\caption{Derived parameters of the target stars.
\label{t_res}} \\
\hline
\tiny{ID} & \tiny{$V$} & \tiny{$(U-V)$} & \tiny{$(B-V)$} & \tiny{$T_\mathrm{eff}$} & \tiny{$log(g)$} &
\tiny{$log(\frac{N(He)}{N(H)})$} & M & \tiny{$E(B-V$)} & \tiny{$RV_H$} & Notes \\
 & & & & K & dex & dex & $M_{\sun}$ & & km~s$^{-1}$ & \\
\hline
\endfirsthead
\caption{continued.}\\
\hline
\tiny{ID} & \tiny{$V$} & \tiny{$(U-V)$} & \tiny{$(B-V)$} & \tiny{T$_\mathrm{eff}$} & \tiny{$log(g)$} &
\tiny{$log(\frac{N(He)}{N(H)})$} & M & \tiny{$E(B-V$)} & \tiny{$RV_H$} & Notes \\
\hline
\endhead
\hline
\endfoot
75369  & 15.285 & $-$0.048 & 0.054    & 10800$\pm$\ \ 130   & 3.36$\pm$0.06 &        $-$       & 0.27$\pm$0.03 & 0.140   & 228 & \\
75469  & 15.437 & $-$0.066 & 0.033    & 11600$\pm$\ \ 120   & 3.76$\pm$0.03 & $-$2.10$\pm$0.54 & 0.52$\pm$0.05 & 0.130   & 236 & \\
76912  & 15.687 & $-$0.295 & 0.001    & 11900$\pm$\ \ 180   & 3.60$\pm$0.03 & $-$2.05$\pm$0.51 & 0.28$\pm$0.03 & 0.106   & 224 & \\
77018  & 15.647 & $-$0.255 & 0.031    & 13000$\pm$\ \ 160   & 3.92$\pm$0.03 & $-$2.05$\pm$0.27 & 0.52$\pm$0.05 & 0.150   & 288 & M \\
77151  & 15.149 & 0.139    & 0.064    & \ 9730$\pm$\ \ \ \ \ 40 & 3.30$\pm$0.06 &        $-$       & 0.33$\pm$0.03 & 0.106   & 214 & \\
77359  & 14.871 & 0.431    & 0.145    & \ 8700$\pm$\ \ \ \ \ 40 & 3.09$\pm$0.06 &        $-$       & 0.35$\pm$0.04 & 0.136   & 288 & \\
77547  & 14.995 & 0.279    & 0.120    & \ 9300$\pm$\ \ \ \ \ 60 & 3.22$\pm$0.03 &        $-$       & 0.35$\pm$0.03 & 0.144   & 223 & \\
78728  & 15.469 & $-$0.209 & 0.023    & 11900$\pm$\ \ 200   & 3.73$\pm$0.06 & $-$2.08$\pm$0.60 & 0.46$\pm$0.06 & 0.125   & 248 & \\
79423  & 15.442 & $-$0.197 & 0.026    & 12100$\pm$\ \ 200   & 3.65$\pm$0.06 & $-$2.56$\pm$0.78 & 0.38$\pm$0.05 & 0.133   & 189 & \\
79535  & 14.730 & 0.482    & 0.163    & \ 8470$\pm$\ \ \ \ \ 60 & 2.97$\pm$0.09 &        $-$       & 0.33$\pm$0.04 & 0.142   & 228 & \\
79548  & 16.260 & $-$0.572 & 0.004    & 15500$\pm$\ \ 200   & 4.18$\pm$0.03 & $-$2.04$\pm$0.15 & 0.39$\pm$0.04 & 0.157   & 222 & \\
81127  & 15.312 & $-$0.059 & 0.064    & 11800$\pm$\ \ 120   & 3.79$\pm$0.03 & $-$1.57$\pm$0.24 & 0.61$\pm$0.06 & 0.164   & 267 & \\
81531  & 18.464 & $-$1.335 & $-$0.150 & 38700$\pm$\ \ 300   & 5.67$\pm$0.06 & $-$1.02$\pm$0.06 & 0.36$\pm$0.04 & 0.146   & 236 & M \\
82039  & 15.824 & $-$0.423 & 0.000    & 13700$\pm$\ \ 180   & 4.00$\pm$0.03 & $-$1.76$\pm$0.18 & 0.48$\pm$0.05 & 0.131   & 291 & \\
82876  & 16.055 & $-$0.429 & 0.016    & 13500$\pm$\ \ 200   & 3.91$\pm$0.03 & $-$2.11$\pm$0.36 & 0.32$\pm$0.03 & 0.145   & 305 & \\
83092  & 14.968 & 0.255    & 0.099    & \ 9180$\pm$\ \ \ \ \ 40 & 3.10$\pm$0.09 &        $-$       & 0.28$\pm$0.04 & 0.122   & 289 & \\
84344  & 14.846 & 0.392    & 0.138    & \ 8700$\pm$\ \ \ \ \ 70 & 3.00$\pm$0.06 &        $-$       & 0.29$\pm$0.03 & 0.133   & 279 & \\
84407  & 14.905 & 0.217    & 0.105    & \ 9650$\pm$\ \ \ \ \ 30 & 3.27$\pm$0.06 &        $-$       & 0.39$\pm$0.04 & 0.144   & 232 & \\
85323  & 17.128 & $-$0.732 & $-$0.025 & 17000$\pm$\ \ 300   & 4.76$\pm$0.12 & $-$2.15$\pm$0.12 & 0.57$\pm$0.09 & 0.132   & 246 & M \\
85415  & 15.033 & 0.241    & 0.106    & \ 9700$\pm$\ \ \ \ \ 50 & 3.31$\pm$0.06 &        $-$       & 0.38$\pm$0.04 & 0.147   & 302 & \\
85568  & 15.029 & 0.274    & 0.100    & \ 9440$\pm$\ \ \ \ \ 50 & 3.26$\pm$0.15 &        $-$       & 0.36$\pm$0.06 & 0.130   & 231 & \\
86663  & 15.202 & 0.153    & 0.099    & \ 9640$\pm$\ \ \ \ \ 50 & 3.19$\pm$0.09 &        $-$       & 0.25$\pm$0.03 & 0.141   & 263 & \\
87776  & 17.806 & $-$1.008 & $-$0.115 & 22400$\pm$\ \ 400   & 5.09$\pm$0.03 & $-$1.87$\pm$0.09 & 0.38$\pm$0.04 & 0.090   & 205 & \\
88234  & 15.815 & $-$0.313 & 0.008    & 12500$\pm$\ \ 140   & 3.81$\pm$0.06 & $-$2.20$\pm$0.39 & 0.36$\pm$0.04 & 0.122   & 224 & \\
89168  & 15.291 & $-$0.119 & 0.029    & 10800$\pm$\ \ 170   & 3.27$\pm$0.09 &        $-$       & 0.22$\pm$0.03 & 0.117   & 222 & \\
89638  & 18.547 & $-$1.306 & $-$0.122 & 36600$\pm$1000      & 5.66$\pm$0.12 & $-$0.21$\pm$0.15 & 0.35$\pm$0.06 & 0.165   & 231 & M \\
90381  & 14.891 & 0.364    & 0.122    & \ 9070$\pm$\ \ \ \ \ 50 & 3.16$\pm$0.18 &        $-$       & 0.36$\pm$0.07 & 0.136   & 249 & \\
91164  & 18.577 & $-$1.330 & $-$0.191 & 33100$\pm$1200      & 5.67$\pm$0.09 & $-$1.08$\pm$0.18 & 0.39$\pm$0.07 & 0.072   & 189 & T \\
91573  & 18.313 & $-$1.011 & $-$0.082 & 27400$\pm$1800      & 5.61$\pm$0.12 & $-$2.69$\pm$0.54 & 0.56$\pm$0.16 & 0.148   & 210 & M \\
91877  & 15.352 & $-$0.029 & 0.049    & 10090$\pm$\ \ \ \ 30  & 3.18$\pm$0.12 &        $-$       & 0.19$\pm$0.03 & 0.116   & 246 & \\
92018  & 18.308 & $-$1.146 & $-$0.130 & 28200$\pm$1300      & 5.53$\pm$0.12 & $-$1.92$\pm$0.15 & 0.44$\pm$0.11 & 0.107   & 206 & T \\
93131  & 17.531 & $-$0.778 & $-$0.079 & 17400$\pm$\ \ 500   & 4.73$\pm$0.12 & $-$1.67$\pm$0.18 & 0.35$\pm$0.06 & 0.083   & 172 & \\
93226  & 16.215 & $-$0.561 & $-$0.019 & 15800$\pm$\ \ 400   & 4.14$\pm$0.09 & $-$2.35$\pm$0.27 & 0.36$\pm$0.06 & 0.137   & 251 & \\
94034  & 14.957 & 0.305    & 0.092    & \ 9050$\pm$\ \ \ \ \ 30 & 3.11$\pm$0.15 &        $-$       & 0.30$\pm$0.05 & 0.106   & 217 & \\
95259  & 15.934 & $-$0.401 & 0.022    & 14000$\pm$\ \ 300   & 3.98$\pm$0.09 & $-$2.66$\pm$0.03 & 0.40$\pm$0.06 & 0.158   & 210 & \\
95987  & 18.364 & $-$1.106 & $-$0.100 & 28400$\pm$1200      & 5.47$\pm$0.09 & $-$2.24$\pm$0.33 & 0.36$\pm$0.08 & 0.139   & 180 & T \\
96597  & 15.531 & $-$0.218 & 0.009    & 12100$\pm$\ \ 170   & 3.59$\pm$0.06 & $-$2.46$\pm$0.99 & 0.30$\pm$0.04 & 0.118   & 220 & \\
97034  & 17.500 & $-$0.827 & $-$0.059 & 23400$\pm$1600      & 4.90$\pm$0.15 & $-$2.62$\pm$0.36 & 0.30$\pm$0.10 & 0.159   & 153 & T \\
97088  & 16.109 & $-$0.478 & 0.006    & 13700$\pm$\ \ 300   & 3.74$\pm$0.09 & $-$2.02$\pm$0.30 & 0.21$\pm$0.03 & 0.140   & 206 & \\
98189  & 18.434 & $-$1.348 & $-$0.194 & 36000$\pm$1000      & 5.89$\pm$0.09 & $-$0.92$\pm$0.15 & 0.66$\pm$0.10 & 0.086   & 209 & T \\
98349  & 15.018 & 0.295    & 0.160    & \ 8820$\pm$\ \ \ \ \ 70 & 3.08$\pm$0.21 &        $-$       & 0.29$\pm$0.07 & 0.160   & 191 & \\
98857  & 18.855 & $-$1.342 & $-$0.224 & 35600$\pm$\ \ 900   & 5.83$\pm$0.12 & $-$0.62$\pm$0.12 & 0.40$\pm$0.07 & 0.053   & 200 & \\
99148  & 14.914 & 0.378    & 0.112    & \ 8540$\pm$\ \ \ \ \ 70 & 2.92$\pm$0.18 &        $-$       & 0.24$\pm$0.05 & 0.100   & 204 & \\
100171 & 18.900 & $-$1.361 & $-$0.162 & 33900$\pm$1700      & 5.73$\pm$0.15 & $-$0.23$\pm$0.15 & 0.32$\pm$0.08 & 0.106   & 230 & M \\
100288 & 15.541 & $-$0.204 & 0.014    & 12000$\pm$\ \ 190   & 3.78$\pm$0.12 & $-$1.72$\pm$0.36 & 0.47$\pm$0.08 & 0.118   & 161 & \\
100817 & 15.190 & $-$0.058 & 0.059    & 12000$\pm$\ \ 110   & 3.71$\pm$0.09 & $-$1.50$\pm$0.30 & 0.56$\pm$0.07 & 0.163   & 226 & \\
101202 & 16.709 & $-$0.625 & 0.026    & 18000$\pm$\ \ 400   & 4.65$\pm$0.09 & $-$2.15$\pm$0.21 & 0.59$\pm$0.07 & 0.199   & 276 & \\
101650 & 15.558 & $-$0.217 & 0.014    & 11800$\pm$\ \ 140   & 3.60$\pm$0.09 & $-$2.07$\pm$0.48 & 0.31$\pm$0.04 & 0.117   & 204 & \\
102372 & 14.908 & 0.339    & 0.110    & \ 8800$\pm$\ \ \ \ \ 80 & 3.01$\pm$0.24 &        $-$       & 0.27$\pm$0.07 & 0.112   & 188 & \\
103232 & 18.561 & $-$1.293 & $-$0.149 & 35200$\pm$\ \ 600   & 5.49$\pm$0.06 & $-$0.83$\pm$0.09 & 0.24$\pm$0.03 & 0.132   & 224 & \\
104153 & 16.090 & $-$0.480 & $-$0.013 & 14200$\pm$\ \ 500   & 4.00$\pm$0.12 & $-$2.63$\pm$0.75 & 0.36$\pm$0.07 & 0.124   & 215 & \\
104974 & 14.773 & 0.444    & 0.149    & \ 8450$\pm$\ \ \ 120  & 3.06$\pm$0.21 &        $-$       & 0.39$\pm$0.09 & 0.121   & 302 & \\
105290 & 19.180 & $-$1.252 & $-$0.035 & 36000$\pm$2000      & 5.33$\pm$0.07 & +0.69$\pm$0.42   & 0.12$\pm$0.04 & $-$      & 237 & \\
105596 & 14.924 & 0.353    & 0.114    & \ 8430$\pm$\ \ \ \ \ 70 & 3.04$\pm$0.09 &        $-$       & 0.33$\pm$0.04 & 0.085   & 218 & \\
107532 & 15.015 & 0.263    & 0.077    & \ 9590$\pm$\ \ \ \ \ 50 & 3.38$\pm$0.15 &        $-$       & 0.46$\pm$0.08 & 0.110   & 166 & \\
106348 & 15.156 & 0.112    & 0.070    & \ 9810$\pm$\ \ \ \ \ 70 & 3.28$\pm$0.09 &        $-$       & 0.31$\pm$0.04 & 0.116   & 203 & \\
108101 & 14.953 & 0.205    & 0.070    & \ 9220$\pm$\ \ \ \ \ 70 & 3.10$\pm$0.12 &        $-$       & 0.28$\pm$0.04 & 0.095   & 201 & \\
108102 & 19.035 & $-$1.333 & $-$0.150 & 37000$\pm$3000      & 5.79$\pm$0.33 & +1.33$\pm$0.24   & 0.29$\pm$0.13 & 0.142   & 233 & \\
108309 & 18.629 & $-$1.298 & $-$0.123 & 38300$\pm$\ \ 900   & 5.85$\pm$0.15 & $-$0.71$\pm$0.15 & 0.47$\pm$0.09 & 0.172   & 228 & \\
109104 & 14.985 & 0.119    & 0.069    & \ 9860$\pm$\ \ \ \ \ 70 & 3.27$\pm$0.18 &        $-$       & 0.35$\pm$0.07 & 0.117   & 288 & \\
109474 & 16.806 & $-$0.672 & $-$0.030 & 17600$\pm$\ \ 400   & 4.40$\pm$0.09 & $-$1.85$\pm$0.12 & 0.31$\pm$0.05 & 0.145   & 225 & \\
113991 & 18.357 & $-$1.502 & $-$0.243 & 64000$\pm$13000     & 5.77$\pm$0.15 & $-$0.09$\pm$0.42 & 0.28$\pm$0.23 & $-$      & 263 & \\
114321 & 16.643 & $-$0.746 & 0.005    & 18200$\pm$\ \ 400   & 4.35$\pm$0.09 & $-$1.66$\pm$0.12 & 0.30$\pm$0.05 & 0.187   & 268 & \\
114375 & 16.888 & $-$0.674 & $-$0.003 & 17400$\pm$\ \ 400   & 4.68$\pm$0.09 & $-$1.86$\pm$0.18 & 0.56$\pm$0.09 & 0.162   & 255 & \\
128044 & 14.923 & 0.258    & 0.082    & \ 8840$\pm$\ \ \ \ \ 30 & 3.12$\pm$0.09 &        $-$       & 0.34$\pm$0.04 & 0.082   & 219 & \\
129495 & 15.018 & 0.358    & 0.345    & \ 7800$\pm$\ \ \ \ \ 50 & 3.04$\pm$0.15 &        $-$       & 0.41$\pm$0.07 & 0.233   & 191 & \\
131429 & 15.435 & $-$0.176 & 0.002    & 10810$\pm$\ \ \ \ 70  & 3.67$\pm$0.09 &        $-$       & 0.48$\pm$0.06 & 0.081   & 304 & \\
132039 & 15.419 & $-$0.130 & 0.011    & 10290$\pm$\ \ \ \ 40  & 3.42$\pm$0.09 &        $-$       & 0.30$\pm$0.04 & 0.080   & 288 & \\
133061 & 15.217 & $-$0.102 & 0.051    & 13600$\pm$\ \ 200   & 3.90$\pm$0.12 & $-$2.75$\pm$0.48 & 0.68$\pm$0.11 & 0.182   & 191 & \\
133073 & 15.033 & 0.176    & 0.127    & \ 8900$\pm$\ \ \ \ \ 80 & 3.15$\pm$0.21 &        $-$       & 0.33$\pm$0.07 & 0.130   & 240 & \\
133686 & 14.796 & 0.192    & 0.102    & \ 8880$\pm$\ \ \ \ \ 20 & 3.11$\pm$0.15 &        $-$       & 0.37$\pm$0.06 & 0.105   & 268 & \\
133674 & 14.994 & 0.203    & 0.096    & \ 8540$\pm$\ \ \ \ \ 60 & 3.07$\pm$0.09 &        $-$       & 0.32$\pm$0.04 & 0.075   & 249 & \\
134857 & 14.858 & 0.295    & 0.108    & \ 8200$\pm$\ \ \ 150  & 3.07$\pm$0.09 &        $-$       & 0.41$\pm$0.06 & 0.054   & 204 & \\
136031 & 15.026 & 0.183    & 0.087    & \ 9090$\pm$\ \ \ \ \ 40 & 3.14$\pm$0.15 &        $-$       & 0.30$\pm$0.05 & 0.103   & 196 & \\
136206 & 14.855 & 0.358    & 0.144    & \ 8410$\pm$\ \ \ \ \ 90 & 3.11$\pm$0.06 &        $-$       & 0.42$\pm$0.05 & 0.108   & 239 & \\
137401 & 15.084 & 0.155    & 0.062    & \ 8980$\pm$\ \ \ \ \ 50 & 3.03$\pm$0.12 &        $-$       & 0.23$\pm$0.03 & 0.076   & 288 & \\
137858 & 14.715 & 0.380    & 0.151    & \ 8100$\pm$\ \ \ 100  & 2.93$\pm$0.12 &        $-$       & 0.35$\pm$0.05 & 0.097   & 200 & \\
131319 & 15.217 & $-$0.022 & 0.044    & \ 9340$\pm$\ \ \ \ \ 60 & 3.14$\pm$0.12 &        $-$       & 0.24$\pm$0.04 & 0.074   & 250 & \\
133035 & 14.899 & 0.139    & 0.074    & \ 9360$\pm$\ \ \ \ \ 50 & 3.20$\pm$0.18 &        $-$       & 0.36$\pm$0.07 & 0.103   & 182 & \\
133511 & 15.296 & $-$0.028 & 0.029    & \ 9890$\pm$\ \ \ \ \ 70 & 3.42$\pm$0.15 &        $-$       & 0.37$\pm$0.06 & 0.073   & 209 & \\
133767 & 15.584 & $-$0.381 & $-$0.021 & 11400$\pm$\ \ 170   & 3.36$\pm$0.12 &        $-$       & 0.19$\pm$0.03 & 0.078   & 226 & \\
134888 & 15.386 & $-$0.142 & 0.026    & 10500$\pm$\ \ 110   & 3.49$\pm$0.18 &        $-$       & 0.35$\pm$0.07 & 0.101   & 160 & \\
135572 & 15.992 & $-$0.559 & $-$0.050 & 13000$\pm$\ \ 300   & 3.87$\pm$0.09 & $-$2.24$\pm$0.54 & 0.33$\pm$0.05 & 0.071   & 230 & M \\
135942 & 15.924 & $-$0.507 & $-$0.023 & 13200$\pm$\ \ 180   & 3.96$\pm$0.09 & $-$2.77$\pm$0.60 & 0.43$\pm$0.06 & 0.099   & 216 & \\
136852 & 15.427 & $-$0.124 & 0.019    & 10340$\pm$\ \ \ \ 50  & 3.40$\pm$0.12 &        $-$       & 0.28$\pm$0.04 & 0.091   & 242 & \\
137998 & 15.686 & $-$0.382 & $-$0.025 & 11300$\pm$\ \ 140   & 3.38$\pm$0.09 &        $-$       & 0.18$\pm$0.02 & 0.072   & 206 & \\
141523 & 15.574 & $-$0.212 & 0.006    & 10900$\pm$\ \ 190   & 3.55$\pm$0.12 &        $-$       & 0.32$\pm$0.05 & 0.091   & 123 & \\
144281 & 15.328 & $-$0.026 & 0.033    & \ 9650$\pm$\ \ \ \ \ 50 & 3.32$\pm$0.18 &        $-$       & 0.30$\pm$0.06 & 0.071   & 248 & \\
154412 & 17.863 & $-$1.422 & $-$0.235 & 36300$\pm$\ \ 200   & 5.30$\pm$0.06 & $-$0.99$\pm$0.09 & 0.29$\pm$0.03 & 0.054   & 232 & \\
156638 & 18.680 & $-$1.504 & $-$0.272 & 47000$\pm$3000      & 5.57$\pm$0.06 & $-$2.24$\pm$0.39 & 0.18$\pm$0.04 & 0.042   & 280 & \\
157531 & 18.277 & $-$0.952 & 0.096    & 43000$\pm$\ \ 800   & 5.46$\pm$0.03 & $-$0.90$\pm$0.18 & 0.23$\pm$0.03 & 0.402   & 223 & \\
166106 & 19.127 & $-$1.431 & $-$0.399 & 35000$\pm$\ \ 800   & 5.59$\pm$0.09 & $-$0.70$\pm$0.12 & 0.18$\pm$0.03 & $-$0.121 & 223 & T \\
167821 & 18.754 & $-$1.434 & $-$0.278 & 37400$\pm$\ \ 500   & 5.72$\pm$0.06 & $-$1.02$\pm$0.06 & 0.32$\pm$0.04 & 0.014   & 247 & \\
172573 & 16.538 & $-$0.722 & $-$0.085 & 15800$\pm$\ \ 300   & 4.24$\pm$0.06 & $-$2.13$\pm$0.18 & 0.34$\pm$0.04 & 0.070   & 260 & \\
173876 & 17.685 & $-$0.955 & $-$0.064 & 23800$\pm$\ \ 500   & 5.04$\pm$0.09 & $-$2.33$\pm$0.15 & 0.34$\pm$0.05 & 0.154   & 262 & T \\
174767 & 16.836 & $-$0.865 & $-$0.134 & 16800$\pm$\ \ 300   & 4.39$\pm$0.03 & $-$1.88$\pm$0.12 & 0.32$\pm$0.04 & 0.031   & 245 & \\
175847 & 18.547 & $-$1.263 & $-$0.235 & 37900$\pm$\ \ 900   & 5.82$\pm$0.12 & $-$0.59$\pm$0.12 & 0.48$\pm$0.08 & 0.059   & 264 & \\
178139 & 18.078 & $-$0.796 & 0.092    & 32300$\pm$\ \ 600   & 5.47$\pm$0.06 & $-$0.96$\pm$0.12 & 0.40$\pm$0.05 & 0.356   & 257 & \\
180700 & 17.444 & $-$0.844 & $-$0.185 & 18000$\pm$\ \ 400   & 4.65$\pm$0.06 & $-$2.15$\pm$0.18 & 0.29$\pm$0.04 & $-$0.012 & 275 & \\
181428 & 18.851 & $-$1.620 & $-$0.281 & 38400$\pm$1600      & 5.85$\pm$0.09 & $-$2.63$\pm$0.33 & 0.38$\pm$0.08 & 0.014   & 308 & \\
183124 & 18.006 & $-$0.717 & 0.100    & 31700$\pm$\ \ 900   & 5.36$\pm$0.06 & $-$1.79$\pm$0.15 & 0.34$\pm$0.05 & 0.366   & 209 & T \\
224916 & 18.530 & $-$1.421 & $-$0.191 & 36300$\pm$\ \ 800   & 5.77$\pm$0.09 & $-$0.50$\pm$0.09 & 0.46$\pm$0.07 & 0.093   & 223 & \\
225063 & 18.692 & $-$1.366 & $-$0.231 & 36200$\pm$\ \ 600   & 5.72$\pm$0.06 & $-$0.74$\pm$0.06 & 0.35$\pm$0.04 & 0.053   & 258 & \\
225931 & 17.101 & $-$0.800 & $-$0.069 & 17200$\pm$\ \ 300   & 4.51$\pm$0.06 & $-$1.65$\pm$0.09 & 0.32$\pm$0.04 & 0.097   & 269 & \\
229084 & 18.171 & $-$1.155 & $-$0.136 & 28600$\pm$\ \ 700   & 5.59$\pm$0.12 & $-$2.19$\pm$0.18 & 0.57$\pm$0.10 & 0.101   & 245 & M,T \\
229880 & 16.801 & $-$0.765 & $-$0.032 & 17900$\pm$\ \ 400   & 4.46$\pm$0.06 & $-$1.81$\pm$0.15 & 0.35$\pm$0.05 & 0.145   & 231 & \\
230786 & 17.993 & $-$1.000 & $-$0.057 & 31800$\pm$\ \ 700   & 5.65$\pm$0.06 & $-$1.50$\pm$0.09 & 0.66$\pm$0.09 & 0.203   & 203 & T \\
232593 & 18.429 & $-$1.364 & $-$0.167 & 36600$\pm$\ \ 900   & 5.73$\pm$0.09 & $-$0.55$\pm$0.12 & 0.45$\pm$0.07 & 0.119   & 214 & \\
232682 & 17.262 & $-$0.948 & $-$0.112 & 22100$\pm$\ \ 500   & 4.87$\pm$0.09 & $-$1.93$\pm$0.09 & 0.39$\pm$0.06 & 0.096   & 205 & T \\
233133 & 17.457 & $-$1.004 & $-$0.100 & 24100$\pm$\ \ 600   & 5.04$\pm$0.06 & $-$2.30$\pm$0.15 & 0.41$\pm$0.06 & 0.121   & 209 & M,T \\
234000 & 16.650 & $-$0.743 & $-$0.067 & 16200$\pm$\ \ 400   & 4.31$\pm$0.09 & $-$1.76$\pm$0.12 & 0.34$\pm$0.06 & 0.091   & 255 & M \\
234333 & 17.802 & $-$1.345 & $-$0.189 & 35300$\pm$\ \ 800   & 5.30$\pm$0.06 & $-$2.35$\pm$0.30 & 0.31$\pm$0.04 & 0.096   & 241 & M \\
235103 & 17.524 & $-$0.964 & $-$0.077 & 23200$\pm$\ \ 600   & 4.74$\pm$0.15 & $-$1.71$\pm$0.12 & 0.21$\pm$0.04 & 0.143   & 236 & T \\
236142 & 17.888 & $-$1.165 & $-$0.149 & 25600$\pm$\ \ 600   & 5.22$\pm$0.06 & $-$3.37$\pm$0.18 & 0.37$\pm$0.05 & 0.076   & 212 & T \\
236428 & 17.609 & $-$0.873 & $-$0.066 & 23100$\pm$\ \ 700   & 5.08$\pm$0.12 & $-$3.26$\pm$0.15 & 0.42$\pm$0.08 & 0.145   & 199 & T \\
\hline
\end{longtable}}

\end{document}